\documentstyle[12pt,epsfig]{article} 
\def\baselinestretch{1.3}
\newcommand{\ba}{\begin{array}}
\newcommand{\ea}{\end{array}}
\newcommand{\bd}{\begin{displaymath}}
\newcommand{\ed}{\end{displaymath}}
\newcommand{\be}{\begin{equation}}
\newcommand{\ee}{\end{equation}}
\newcommand{\bea}{\begin{eqnarray}}
\newcommand{\eea}{\end{eqnarray}}
\newcommand{\sla}[1]{/\!\!\!#1}

\def\a{\alpha}

\def\m{\mu}
\def\n{\nu}

\def\q2 {q^2}

\def\td {\tilde}
\def\r {\rightarrow}

\def\miss {\hspace{-0.5cm}\slash~~}

\def\rslep {\tilde{e_R}}
\def\rsnu {\tilde{\nu}_R}
\def\snu {\tilde{\nu}}
\def\lslep {\tilde{e_L}}
\def\stau {\tilde{\tau}}
\def\mer {m_{\rslep}}
\def\mmr {m_{\tilde{\mu}_R}}
\def\mml {m_{\tilde{\mu}_L}}
\def\mel {m_{\lslep}}

\def\bt{\begin{table}}
\def\et{\end{table}}

\catcode`@=11 
\def \gsim{\mathrel{\mathpalette\@versim>}}
\def \lsim{\mathrel{\mathpalette\@versim<}}
\def \@versim#1#2{\lower0.4ex\vbox{\baselineskip\z@skip\lineskip\z@skip
     \lineskiplimit\z@\ialign{$\m@th#1\hfil##\hfil$%
     \crcr#2\crcr\sim\crcr}}}
\catcode`@=12 

\begin{document}

\begin{flushright}
{RECAPP-HRI-2009-004}
\end{flushright}

\begin{center}

{\large\bf Neutralino reconstruction in supersymmetry with long-lived staus}\\[15mm]
Sanjoy Biswas\footnote{E-mail: sbiswas@mri.ernet.in}
and Biswarup Mukhopadhyaya\footnote{E-mail: biswarup@mri.ernet.in}\\
{\em Regional Centre for Accelerator-based Particle Physics \\
     Harish-Chandra Research Institute\\
Chhatnag Road, Jhunsi, Allahabad - 211 019, India}
\\[20mm] 
\end{center}

\begin{abstract} 
  We consider a supergravity (SUGRA) scenario, with universal scalar
  and gaugino masses at high scale, with a right-chiral neutrino
  superfield included in the spectrum. Such a scenario can have a
  lightest supersymmetric particle (LSP) dominated by the right
  sneutrino and a stau as the next-to lightest supersymmetric
  particle (NLSP). Since decays of all particles into the LSP are
  suppressed by the neutrino Yukawa coupling, the signal of
  supersymmetry consists in charged tracks of stable particles in the
  muon chamber. We demonstrate how a neutralino decaying into a tau
  and the stau-NLSP can be fully reconstructed over substantial areas
  in the SUGRA parameter space. We also suggest event selection
  criteria for eliminating backgrounds, including combinatorial ones,
  and use a new method for the extraction of the mass of the
  stau-NLSP, using its three-momentum as obtained from the curvature of
  the charged track.

\end{abstract}

\vskip 1 true cm

\newpage
\setcounter{footnote}{0}

\def\baselinestretch{1.5}
\section{Introduction}

Searches for supersymmetry (SUSY) \cite{Book,Sally,mSUGRA} at the Large
Hadron Collider (LHC) are largely based on signals with missing
transverse energy ($\sla{E_T}$).  This is because SUSY, in its
${\mathcal R}$-parity conserving form (with ${\mathcal R}$-parity defined by 
${\mathcal R} =(-)^{3B+L+2S}$), offers the lightest supersymmetric particle (LSP)
which is stable, and, if electrically uncharged and weakly
interacting, is potentially a cold dark matter candidate. The lightest
neutralino ($\chi^0_1$) turns out to be the LSP in most theoretical
models. All SUSY cascades at collider experiments should culminate in
the pair-production of the LSP within the detector itself. The
neutral and non-strongly interacting character of the LSP results in
its invisibility at colliders, and thus a large energy-momentum
imbalance, together with energetic jets and/or leptons emerges as the
characteristic signal of SUSY containing a dark matter candidate.

It should be remembered, though, that the above possibility is not
unique. Apart from the lightest neutralino, the
left-chiral sneutrinos in the minimal SUSY standard model (MSSM) can
in principle be a dark matter candidate as well. This is, however,
strongly disfavoured by direct dark matter search experiments, because
the SU(2) interaction of a left-chiral sneutrino (as opposed to the
U(1) or Yukawa dominated interaction of a neutralino) gives rise to
unacceptably large cross-sections of elastic scattering with dark
matter detectors. In addition, a left-chiral sneutrino LSP is
difficult to accommodate in a scenario where the SUSY breaking masses
evolve from `universal' scalar and gaugino mass parameters at a high
scale \cite{P.Nath:2008hs}.

The situation changes if one has right-chiral neutrino superfields in
addition, a possibility that often haunts us as evidence piles up in
favour of neutrino masses and mixing
\cite{Mohapatra:2005wg,nudata}. It has been shown in some recent works
\cite{Asaka:2005cn}
that such a right chiral sneutrino may pass off as a dark matter
candidate without any contradiction from available data \cite{wmap}. Since the
right-chiral sneutrino has no gauge interaction, the only way it can
interact with matter is via neutrino Yukawa coupling, the strength of
its interaction is too feeble to be seen in dark matter search
experiments. In such a
case, the next-to-lightest SUSY particle (NLSP), too, has an
excruciatingly slow rate of decay into the LSP dominated by
right-chiral sneutrino states. Consequently, the NLSP is stable on the
scale of collider detectors, and, in cases where it is a charged
particle, the essence of the SUSY signal lies not in $\sla{E_T}$ but in
charged tracks due to massive particles, seen in the muon chambers.

Such stable charged particles can in principle be distinguished
from muons through a number of techniques. These include the
measurement of time delay between the inner tracking chamber
and the muon chamber, the degree of ionisation, and also more
exotic proposals such as the absorption of the stable particles 
in a chamber which can be subsequently emptied underground
to observe the decays \cite{Hamaguchi:2006vu}.
While these are all of sufficient importance and
interest, some of us have shown in earlier works  
\cite{Gupta:2007rc,Choudhury:2008gb} that there are some
very good kinematic discriminators for such stable charged particles,
which make the signals practically background-free for both stau and stop
NLSP. Event selection criteria based on the transverse momentum $p_T$ 
of the tracks, in conjuction
with quantities such as the scalar sum of all visible transverse momenta
and the invariant mass of track pairs, are found to be useful 
in this respect. In this work, we perform a detailed simulation
of signals, backgrounds and mistags to show that the masses of neutralinos
can be reconstructed to a high level of precision for a scenario with
$\stau$ NLSP and an LSP dominated by the right-chiral sneutrino of the
third family. We use the technique of tau reconstruction for this purpose.
Also, we depend on neither ionisation nor time delay for extracting
the mass of the stable stau, but rather obtain
it using an algorithm that depends on event-by-event information
on two taus and two stable tracks in the final state.

It should be mentioned that the signal discussed here as well as the
reconstruction technique advocated by us is not limited to scenarios
with right-sneutrino LSP alone. One can have stable staus in SUSY,
when, for example, one has a gravitino  LSP in a supergravity (SUGRA) model
\cite{gravitinoLSP}.
They can be envisioned in gauge-mediated SUSY breaking theories as well 
\cite{gmsbnlsp}.
In the MSSM, too, one can have the
so-called co-annihilation region of dark matter, where a stau and
the neutralino LSP are closely degenerate, leading to a quasi-stable
character of the former \cite{coanni}. It should be emphasized that our 
suggested procedure is applicable to all of these cases.
What we find as a bonus is that scenarios with stau NLSP and
right-sneutrino LSP occur rather naturally in a SUGRA model 
\cite{Asaka:2005cn,Gupta:2007rc}
with a universal scalar mass which is the origin of the right-sneutrino
mass as well. Thus the mere addition of a right-sneutrino superfield, 
which is perhaps the most minimal input to explain neutrino masses,
can turn a mSUGRA theory into one with a stau NLSP and a sneutrino
LSP. Thus one can identify regions in the $m_{0} - M_{1/2}$ plane
of the theory, where the reconstruction of unstable neutralinos
is feasible at the LHC.

In section 2, we discuss the scenario under investigation as well as
the super particle spectrum, and motivate the choice of benchmark
points used for demonstrating our claims, in the context of a
supergravity scenario.  The signal looked for, the corresponding
standard model backgrounds and the event selection criteria chosen by
us are discussed in section 3.  Section 4 contains discussions on the
various steps in reconstructing neutralinos. The regions in the $m_{0}
- M_{1/2}$ plane in a SUGRA scenario, where neutralino reconstruction
is possible in our method, are also pointed out in this section.  We
summarise and conclude in section 5.

\section{Right sneutrino LSP in supergravity}
The superpotential of the minimal SUSY Standard Model (MSSM)
\cite{S.P.Martin1} is given (suppressing family indices) 
by \be W_{MSSM} = y_l
\hat{L}\hat{H_d}\hat{E^c} + y_d \hat{Q}\hat{H_d}\hat{D^c}+y_u \hat{Q}
\hat{H_u} \hat{U^c}+\mu\hat{H_d}\hat{H_u} \ee where $\hat{H_d}$ and
$\hat{H_u}$, respectively, are the Higgs superfields that give mass
respectively to the $T_{3}=-1/2$ and $T_{3}=+1/2$ fermions. $y's$
are the strengths of Yukawa interactions. $\hat{L}$ and $\hat{Q}$ are
the left-handed lepton and quark superfields respectively, whereas
$\hat{E^c}$, $\hat{D^c}$ and $\hat{U^c}$, in that order, are the right
handed gauge singlet charged lepton, down-type and up-type quark
superfields. $\mu$ is the Higgsino mass parameter.

As has been already mentioned, the MSSM must be additionally equipped 
to explain non-vanishing neutrino masses. 
Phenomenologically, the simplest (though not  theoretically the most 
satisfying) way to do so is to assume neutrinos to be of Dirac type
and simply add one right-handed neutrino superfield to each family.
The superpotential of the minimal SUSY standard model 
is thus extended by just one term per family, of the form
\bea
W_\nu^R=y_\nu \hat{H}_u \hat{L}\hat{\nu}^c_R
\eea
However, having such small Dirac masses for the neutrinos would imply
that the neutrino Yukawa couplings ($y_\nu$) are quite small ($\sim 10^{-13}$).
The above term in the superpotential obviously implies the inclusion of
right-handed sneutrinos in the (super)particle spectrum, and these
sneutrinos will have all their 
interactions proportional to the corresponding neutrino masses. Thus
the dominantly right-handed eigenstate of the tau-sneutrino might become 
a viable dark matter candidate, without coming into conflict with
dark matter search limits, thanks to its extremely feeble strength
of interaction with all matter.

Interestingly, scenarios where the MSSM is embedded in
a bigger, high-scale framework for SUSY breaking can support the 
above situation. The most commonly invoked scheme is 
minimal supergravity (mSUGRA) where all scalar (gaugino)
masses at low-energy arise from a universal mass
parameter $m_0(M_{1/2})$ at a high scale where supergravity,
or local SUSY, is broken. If one adds a right-chiral neutrino superfield,
then the right-sneutrino mass  may be assumed to originate in the
same parameter $m_0$. As we shall see below, this causes the
physical state dominated by the right-chiral sneutrino to become the LSP.
It has been shown that such a possibility is
consistent with all experimental bounds \cite{Amsler:2008zzb} 
and also compatible with the dark matter density in the 
Universe \cite{Asaka:2005cn,wmap}. 

The neutrinos masses can be schematically shown as
\bea
m_\nu = y_\nu \left<H_u^0\right> = y_\nu v~\sin\beta
\eea

\noindent
where $\tan\beta$ is the ratio of the vacuum expectation values
of the two Higgs doublets that give masses to the up-and
down-type quarks respectively.
The actual mass eigenvalues will of course depend on the Yukawa
coupling matrix. This, however, 
does not affect the collider signals of the SUSY scenario under consideration 
here, as the interaction strengths of the dominantly right-chiral
states are always very small in magnitude.

Upon inclusion of right-chiral neutrino superfield into the SUGRA
framework, the superparticle spectrum mimics the mSUGRA spectrum in
all details except for the identity of the LSP. As has been already
mentioned, SUSY breaking in the hidden sector at high scale is
manifested in universal soft masses for scalars ($m_0$) and gauginos
($M_{1/2}$), together with the trilinear (A) and bilinear (B) SUSY
breaking parameters in the scalar sector (of which the latter is
determined by electroweak symmetry breaking conditions).  Masses for
squarks, sleptons and gauginos, all the mass parameters in the Higgs
sector as well as the Higgsino mass parameter $\mu$ (up to a sign) are
determined, once the high scale of SUSY breaking in the hidden sector
(${\mathcal O}\sim10^{11}$ GeV) are specified. Neglecting inter-family
mixing, the mass terms for sneutrinos arising in this manner are given
by \bea -{\mathcal L}_{soft} \sim {M}_{{\rsnu}}^2
|\widetilde{\nu}_R|^2 + (y_{\nu}{A}_\nu H_u.\widetilde{L}
\widetilde{\nu}_R^c ~~ + ~~h.c.)  \eea where $A_\nu$ is the term
driving left-right mixing in the scalar mass matrix, and is obtained
by running of the trilinear soft SUSY breaking term $A$
\cite{S.P.Martin2}.  The Yukawa couplings can cause large splitting in
the third-family squark and sleptons masses while the first two
families are more closely degenerate. On the other hand, the degree of
left-right mixing of sneutrinos, driven largely by the Yukawa
couplings, is extremely small.

The sneutrino mass-squared matrix is thus of the form
\bea
m_{\tilde{\nu}}^2 = \left ( \begin{array}{cc} {M}_{\tilde{L}}^2 +
\frac{1}{2}m_Z^2\cos 2\beta & y_\nu v({A}_\nu \sin\beta-\mu\cos\beta)\\
y_\nu v({A}_\nu \sin\beta-\mu \cos\beta) & 
{M}_{\tilde{\nu}_R}^2 \end{array} \right)
\eea
where ${M}_{\tilde{L}}$ is the soft scalar mass for the 
left-handed sleptons whereas the ${M}_{\tilde{\nu}_R}$ is 
that for the right-handed sneutrino. In general, 
 ${M}_{\tilde{L}} \ne {M}_{\tilde{\nu}_R}$ because of their different
evolution patterns. In addition, the $D$-term contribution for the 
former causes a difference between the two diagonal entries. 
While the evolution of all other parameters
in this scenario are practically the same as in the MSSM, the
right-chiral sneutrino mass parameter evolves \cite{arkani} at 
the one-loop level as:
\bea
\frac{dM^2_{\rsnu}}{dt} = \frac{2}{16\pi^2}y^2_\nu~A^2_\nu ~~.
\eea

Clearly, the extremely small Yukawa couplings cause  ${M}_{\tilde{\nu}_R}$
to remain nearly frozen at the value $m_0$, whereas the other sfermion
masses are enhanced at the electroweak scale. Thus, for a wide range of
values of the gaugino mass, one naturally has sneutrino LSP, which,
for every family, is dominated by the right-chiral state:
\bea
\tilde{\nu}_1 = - \tilde{\nu}_L \sin\theta + \tilde{\nu}_R \cos\theta
\eea
The mixing angle $\theta$ is given as 
\bea
\tan 2\theta = \frac{2 y_\nu v\sin\beta |\cot\beta\mu -
A_\nu|}{m^2_{\tilde{\nu}_L}-m^2_{\rsnu}}
\eea

\noindent
{which is suppressed by $y_\nu$, especially if the neutrinos
have Dirac masses only. It is to be noted
that all three (dominantly) right sneutrinos have a similar fate
here, and one has near-degeneracy of three such LSPs. However,
of the three charged slepton families, the amount of left-right
mixing is always the largest in the third (being, of course,
more pronounced for large $\tan\beta$), and the lighter stau
(${\stau}_1$) often turns out to be the NLSP in such a scenario. 
\footnote{We have neglected inter-family mixing in the sneutrino 
sector in this study. While near-degenerate 
physical states makes such mixing likely, the
degree of such mixing is model-dependent, and does not generally
affect the fact that all cascades culminate in the lighter stau, 
so long as the latter is the NLSP, which is the scenario studied 
here.}}

Thus the mSUGRA parameter set ($m_0,M_{1/2},A_0,sign(\mu)~{\rm and}~\tan\beta$)
in an ${\mathcal R}$-parity conserving scenario can eminently lead to a spectrum where
all three generations of right-sneutrinos will be either stable 
or metastable but very long-lived, and can lead to different decay chains
of supersymmetric particles, as compared to those with
a neutralino LSP. However, as we shall see below, the deciding
factor is the lighter sneutrino mass eigenstate of the third family,
so long as the state ${\stau}_1$ is the lightest among
the charged sleptons.

All superparticles will have to decay into the lighter sneutrino of 
a particular family via either gauge interactions (such as
$\tilde{\tau}_L \longrightarrow W \tilde{\nu^{\tau}}_1$)  or Yukawa coupling
(such as $\tilde{l}_L \longrightarrow H^- \tilde{\nu}_1$ or
 $\tilde{\nu}_2 \longrightarrow h^0 \tilde{\nu}_1$). In the former case,
the decay depends entirely on the $\tilde{\nu}_L$ content
of  $\tilde{\nu}_1$, which again depends on the neutrino Yukawa coupling.
The same parameter explicitly controls the decay in the latter case, too.
Therefore, while the lighter sneutrinos of the first two families can in 
principle be produced from decays of the corresponding charged 
sleptons, such decays will be always suppressed compared to even 
three-body decays such as 
$\tilde{e}_1 (\tilde{\mu}_1) \longrightarrow e(\mu) \bar{\tau} {\stau}_1$
(when the sleptons are lighter than all neutralinos).
For the $\stau$-NLSP, however, the only available decay channel
is $\stau_1 \longrightarrow W (H^{-})\tilde{\nu}_1$, with either real or
virtual charged Higgs. Both of these decay channels are driven
by the extremely small neutrino Yukawa coupling.

This causes the NLSP to be a long-lived particle and opens
up a whole set of new possibilities for collider signatures for such
long-lived particles, while retaining contributions to 
dark matter from the sneutrino LSP. The NLSP appears stable in collider
detectors and gives highly ionizing charged tracks. 

Apart from a stau, the NLSP could be a chargino, a stop  or a sbottom. 
The former is in general difficult to achieve in a scenario
where the chargino and neutralino masses are determined by the 
same set of electroweak gaugino and Higgsino masses. The phenomenology
of the long-lived stop NLSP \cite{Chou:1999zb}, the likelihood of the corresponding
signals being available at the early phase of the LHC, and the 
potential for the reconstruction of gluino masses have been discussed
in an earlier work \cite{Choudhury:2008gb}. The stau NLSP, as we shall see below, offers a
new opportunity to reconstruct both the lightest and second lightest 
neutralino masses.

It may be noted  here that the 
region of the mSUGRA parameter space where we work 
is consistent with all experimental bounds, including both collider 
and low-energy constraints (such as the LEP and Tevatron constraints 
on the masses of Higgs, gluinos, charginos and so on as well as those from 
$b \rightarrow s\gamma$, correction to 
the $\rho$-parameter, ($g_\mu - 2$) etc.). Our choice of parameters in the
$m_0 - M_{1/2}$ plane would correspond to a stau-LSP without the
right-sneutrino in the (super) particle spectrum.  Such a situation 
would have been ruled out, had not the existence of the right-chiral
neutrino superfield, with the right sneutrino at the bottom of the 
spectrum, been assumed \cite{constraints}.
However with the right-sneutrino as the
LSP, we find this choice to be a preferable and well-motivated option.
The $\snu_1$-LSP arising out of such a choice becomes a viable dark 
matter candidate, though not necessarily the only one. Using the formulae 
given by Moroi {\it et al.} in reference \cite{Asaka:2005cn}, the contribution 
to the relic density ($\Omega h^2$) is found to be about one order of magnitude 
below the acceptable value \cite{wmap}. While this leaves room 
for additional sources of dark matter, the scenario presented here is consistent 
from the viewpoint of over-closure of the Universe.

We focus on both the regions where (a) $m_{{\stau}_1} > m_{{\tilde \nu}_1} + m_W$,
and (b) the above inequality is not satisfied. In the first case, 
the dominant decay mode is the 
two-body decay of the NLSP, ${\stau}_1 \to \tilde{\nu}_1  W$, and, in the second,
the decay takes place via a virtual $W$. However, the decay takes place
outside the detector in both cases. Decays into a charged Higgs 
constitutes a subdominant channel for the lighter stau.
%

Furthermore, we try to identify regions of the parameter space,
where neutralinos decaying into a tau and a stau can be reconstructed,
through the reconstruction of the tau and the detection of the stau in the
muon chamber. The rates for electroweak production of neutralinos
are generally rather low for this process. Therefore, the procedure works better
when neutralinos are produced from the cascade decays of squarks and gluinos.
This is in spite of the additional number of jets in such processes, which
may fake the tau in certain cases and complicate the analysis of the signals.
We are thus limited to those regions of the parameter space, where the gluino 
and squark production rates are appreciable, and therefore
the value of $M_{1/2}$ is not too high.

With all the above considerations in mind, 
we concentrate on the lighter {\it stau} (${\stau}_1$)
to be the NLSP with lifetime large enough to penetrate collider
detectors like the muons themselves. Using the spectrum generator
of ISAJET 7.78 \cite{isajet}, we find that a large
mSUGRA parameter space can realize this scenario of a right-sneutrino
LSP and stau NLSP, provided that $m_0 < m_{1/2}$ and one has 
$tan\beta$ in the range $\gsim$ 25, the latter condition being
responsible for a larger left-right off-diagonal term in the 
stau mass matrix (and thus one smaller eigenvalue). In Table 1 we 
identify a few benchmark points, all within a SUGRA
scenario with universal scalar and gaugino masses, 
characterised by long-lived staus at the LHC.

\begin{table}[htb]
\begin{center}
\begin{tabular}{||c||c|c|c|c|c|c||}
\hline
\hline
{\bf Input}& {\bf BP-1}&{\bf BP-2}&{\bf BP-3}&{\bf BP-4}&{\bf BP-5}&{\bf BP-6} \\
\hline
            &$m_0=100$&$m_0=100$&$m_0=100$&$m_0=100$&$m_0=100$
            &$m_0=100$\\
mSUGRA      &$m_{1/2}=600$&$m_{1/2}=500$&$m_{1/2}=400$&$m_{1/2}=350$
            &$m_{1/2}=325$&$m_{1/2}=325$\\
            &$\tan\beta=30$ &$\tan\beta=30$&$\tan\beta=30$&$\tan\beta=30$&$\tan\beta=30$
&$\tan\beta=25$\\
\hline
$\mel,\mml$   &418&355&292&262&247&247\\
$\mer,\mmr$   &246&214&183&169&162&162\\
$m_{\snu_{eL}},m_{\snu_{\mu L}}$&408&343&279&247&232&232\\
$m_{\snu_{\tau L}}$&395&333&270&239&224&226\\
$m_{\snu_{iR}}$&100&100&100&100&100&100\\
$m_{\stau_1}$&189&158&127&112&106&124\\
$m_{\stau_2}$&419&359&301&273&259&255\\
\hline
$m_{\chi^0_1}$&248&204&161&140&129&129\\
$m_{\chi^0_2}$&469&386&303&261&241&240\\
$m_{\chi^{\pm}_1}$&470&387&303&262&241&241\\
$m_{\tilde{g}}$&1362&1151&937&829&774&774\\
$m_{\tilde{t}_1}$&969&816&772&582&634&543\\
$m_{\tilde{t}_2}$&1179&1008&818&750&683&709\\
$m_{h^0}$ &115&114&112&111&111&111\\
\hline
\hline
\end{tabular}\\
\caption {\small \it Proposed benchmark points (BP) for the study of the stau-NLSP scenario
in the SUGRA  with right-sneutrino LSP. The value of $m_0$ and $M_{1/2}$ are 
given in  GeV. We have also set $A_0=100~GeV$ and $sgn(\mu)=+$ for benchmark 
points under study.}
\label{tab:1}       
\end{center}
\end{table}
In the next section we use these benchmark points to 
discuss the signatures of the stau NLSP at the
LHC and look for the final states in which it is possible 
to reconstruct the neutralinos. 

\section{Signal and backgrounds}

The signal which we have studied as a signature of stau NLSP and motivated by the
possible reconstruction of the neutralinos from the final state, is given by 

\begin{itemize}
\item $2\tau_j+2\stau (charged-track)+E_{T}\miss+X$
\end{itemize}

\noindent
where $\tau_j$ denotes a jet out of a one-prong decay of the tau, $E_{T}\miss$
stands for missing transverse energy and all accompanying hard jets arising 
from cascades are included in X.

We have simulated pp collisions with a centre-of-mass energy $E_{CM}=14 TeV$.  
The prediction of events assumes an integrated luminosity of 
300 $fb^{-1}$. The signal and various backgrounds are 
calculated using PYTHIA (version-6.4.16) \cite{PYTHIA}. 
Numerical values of various parameters, used in our calculation, are 
as follows \cite{Amsler:2008zzb}:
\\

$~~~~~$ $M_Z=91.2$ GeV   $M_W=80.4$ GeV   $M_t=171.4$ GeV  $M_H=120$ GeV\\
$~~~~~~~$ $\a^{-1}_{em}(M_Z)=127.9$ $~~$  $\a_{s}(M_Z)=0.118$\\

We have worked with the {\bf CTEQ5L} parton distribution function 
\cite{Lai:1999wy}. The factorisation and renormalisation scale are set at 
{$\bf \mu_F=\mu_R= m^{final}_{average}$}. In order to make our estimate
conservative, the signal rates have not been multiplied by any
K-factor \cite{ksusy}, while the main background,
namely, that from $t{\bar t}$ production, has been multiplied by a
K-factor of 1.8 \cite{ktt}. The effects of initial state radiation (ISR) and 
final state radiation (FSR) have been considered in our study.

\subsection {Signal subprocesses:}

We have studied all SUSY subprocess leading to the desired final states. 
Neutralinos are mostly produced in cascade decays of strongly 
interacting sparticles. The dominant contributions thus come from
 
\begin{itemize}
\item {\bf gluino pair production:} ${\bf pp\r \tilde{g}\tilde{g}}$

\item {\bf squark pair production:} ${\bf pp\r \td{q_i}\td{q_j}, \td{q_i}\td{q^*_i}}$

\item {\bf associated squark-gluino production:} ${\bf pp\r \tilde{q}\tilde{g}}$
\end{itemize}
\begin{table}[htb]
\begin{center}
\begin{tabular}{||c||c|c|c|c|c|c||}
\hline
\hline
{\bf Subprocesses}& {\bf BP-1}&{\bf BP-2}&{\bf BP-3}&{\bf BP-4}&{\bf BP-5}&{\bf BP-6} \\
\hline
 All SUSY         &1765 &4143 &11726 &20889 &28864 &15439 \\
\hline
 $\td{q}\td{q^*},\td{g}\td{g}$,$\td{q}\td{g}$&1616 &3897 &11061 &19765 &27785 &14426 \\
\hline
\hline
\end{tabular}\\
\caption {\small \it {The number of $2\tau_j+2\stau$ (charged-track)+$E_{T}\miss+X$ events,
satisfying our basic cuts, for $\int{Ldt}=300~GeV$, from various channel.}}
\label{tab:1}       
\end{center}
\end{table}

In addition, electroweak pair-production of neutralinos can also
contribute to the signal we are looking for. The rates are, however,
much smaller [see Table 2]. Moreover, the relatively small masses of 
the lightest
and the second lightest neutralinos (as compared to the squarks and
the gluino) cause the signal from such subprocesses to be drastically
reduced by the cut employed by us on the scalar sum of all visible
$p_T$'s. For example, for benchmark point 1 (BP1), one has less than
5\% of the total contribution from electroweak processes.

The production of squarks and gluinos have potentially large cross
sections at the LHC. For all our benchmark points listed in Table 1,
the gluino is heavier than the squarks.  Thus its dominant decay is
into a squark and a quark. $\chi^0_1$ being mostly $\tilde{B}$
dominated the main contribution to $\chi^0_1$ production comes from
the decay of right handed squarks ($\tilde{q_{R}}$) and its decay
branching ratio into the $\stau$-$\tau$ pair is almost 100\% when it is
just above the lighter stau in the spectrum. On the other hand, the
$\chi^0_2$ is mostly $\td{W_3}$-dominated, and therefore the main
source of its production is cascade decay of left-chiral squarks
($\td{q_L}$). Such a $\chi^0_2$ can also decay into a $\stau$-$\tau$
pair.

If one can obtain complete information on the four-momentum of
the $\stau$ and the $\tau$, it is thus possible to 
reconstruct both $\chi^0_1$ and $\chi^0_2$ using the final state mentioned above. 
The other two heavier neutralinos ($\chi^0_3$ and $\chi^0_4$), due to 
their low production rate and small decay branching ratios into 
the $\stau$-$\tau$ pair, are relatively difficult to reconstruct.

\subsection{Background subprocesses:}

The standard model background to  $2\tau_j+2\stau+E_{T}\miss$ $~$
comes mainly from the following subprocesses :

\begin{itemize}

\item {\bf $t\bar{t}$ production,} ${\bf t\bar{t}\r bWbW}$:\\
Where two of the resulting jets can be faked as a tau-jet, and muons can come from
the W's. One can also have a situation in which 
any (or both) of the b-quark decays semileptonically ($b\r
c\m\n_{\m}$) 
and any (or both) of the 
W decays to a $\tau-\n_{\tau}$ pair. Though the efficiency of a 
non-tau jet being identified as a
narrow tau-like jet is small (as will be discussed in a later section), 
and it is very unlikely
to have isolated muons from semileptonic decays of the b, 
the overwhelmingly large number of $t\bar{t}$ 
events produced at the LHC makes this subprocess quite a serious source 
of backgrounds.

\item {\bf $Z^0$-pair production,} ${\bf ZZ\r 2\tau+2\mu}$:\\
This subprocess also gives an additional contribution 
to the background, when one Z decays into a 
$\tau\tau$ pair and the other one into a pair of muons.  

\item {\bf Associated ZH-production,} ${\bf ZH\r 2\tau+2\mu}$:\\
This subprocess, though have a small cross-section, can contribute to
the background through the decay of $H\r \tau\tau$ and $Z\r \mu\mu$,
which can fake our signal as well.





\end{itemize}

The additional backgrounds from $t\bar{t}W$, $t\bar{t}Z$ and $Z$+jets can be 
suppressed by the same cuts as those described below. Also a higher non-tau-jet
rejection factor and Z invariant mass cut can reduce the $t\bar{t}Z$ and $Z$+jets 
backgrounds considerably.



\subsection {Event selection criteria :}

In selecting the candidate events selected for neutralino
reconstruction, we choose
the two highest-$p_T$ isolated charged tracks showing
up in the muon chamber, both  
with $p_T>~100~GeV$, as stable staus. (See the detailed discussion
later in this section.)
The isolation criteria for the tracks are shown in Table 3. 
In addition to the $p_T$ cut, the scalar sum of all jets and
lepton in each event is required to be greater than 1 TeV.
It is clear from Figures 1 and 2 that the standard
model backgrounds are effectively eliminated through 
the above criteria. In addition, we require
two $\tau$-jets with $p_T>~50~GeV$, and 
$\sla{E_T}>~40~GeV$ for each event. The justification for both
of these cuts is provided when we discuss $\tau$-tagging and 
reconstruction.

The identification and 3-momentum reconstruction of the
charged track at the muon chamber is done following 
the same criteria and procedure as those for muons.

\newpage
\begin{figure}[htbp]
\begin{center}
\centerline{\epsfig{file=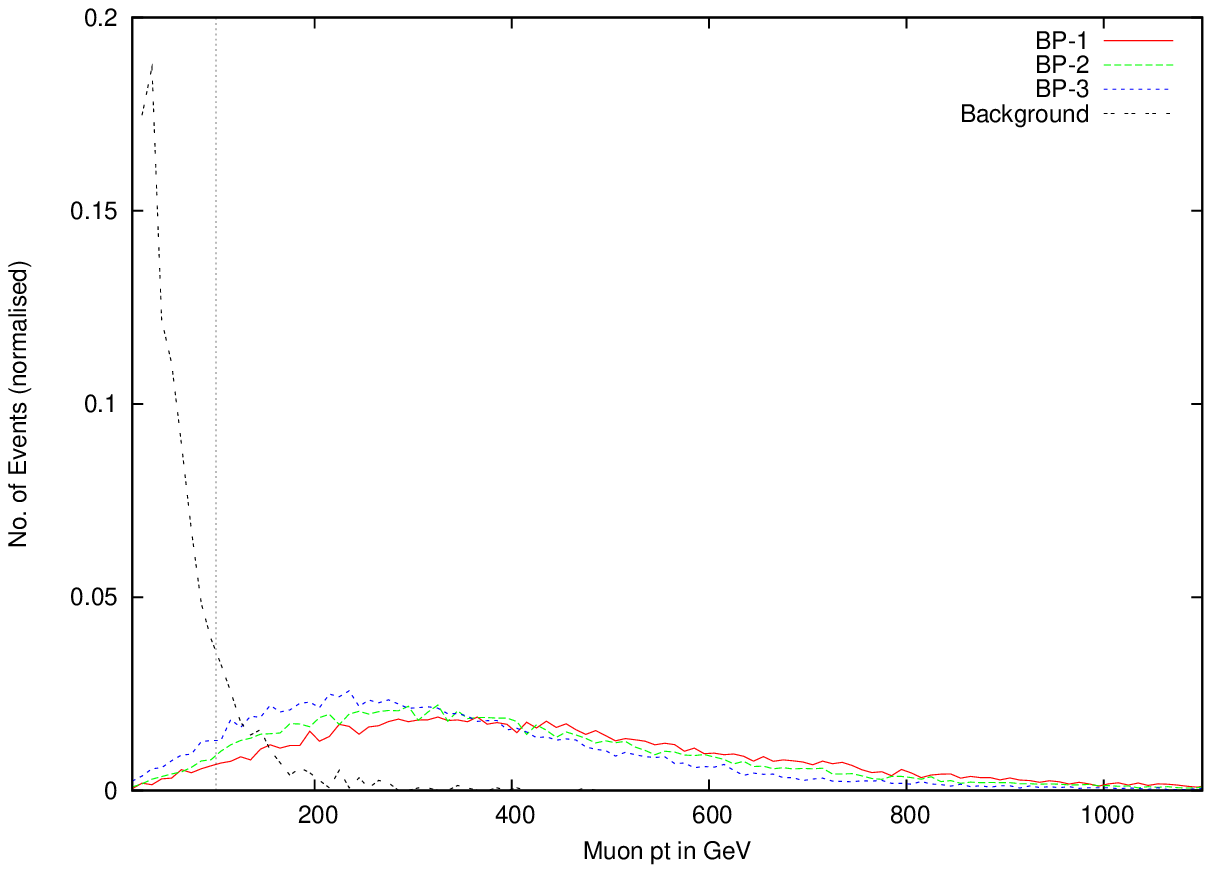,width=7.0cm,height=6.0cm,angle=-0}
\hskip 20pt \epsfig{file=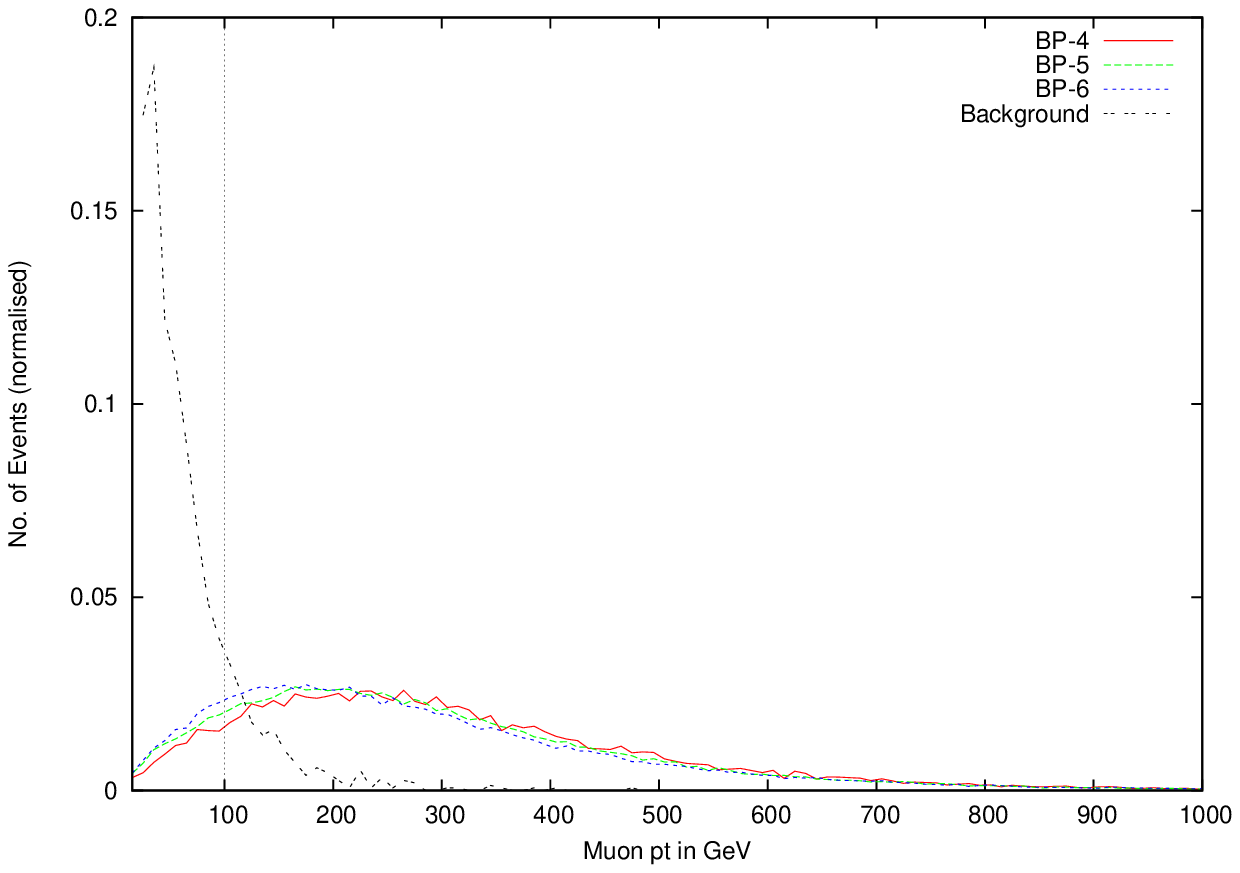,width=7.0cm,height=6.0cm,angle=-0}}
\caption{$p_T$ distribution (normalised to unity) of the muon like track 
for the signal and the background, 
for all benchmark points. The vertical lines indicate the effects of a $p_T$ cut
at 100 GeV.} 
\end{center}
\end{figure}

\vspace{-0.5cm}
\begin{figure}[htbp]
\begin{center}
\centerline{\epsfig{file=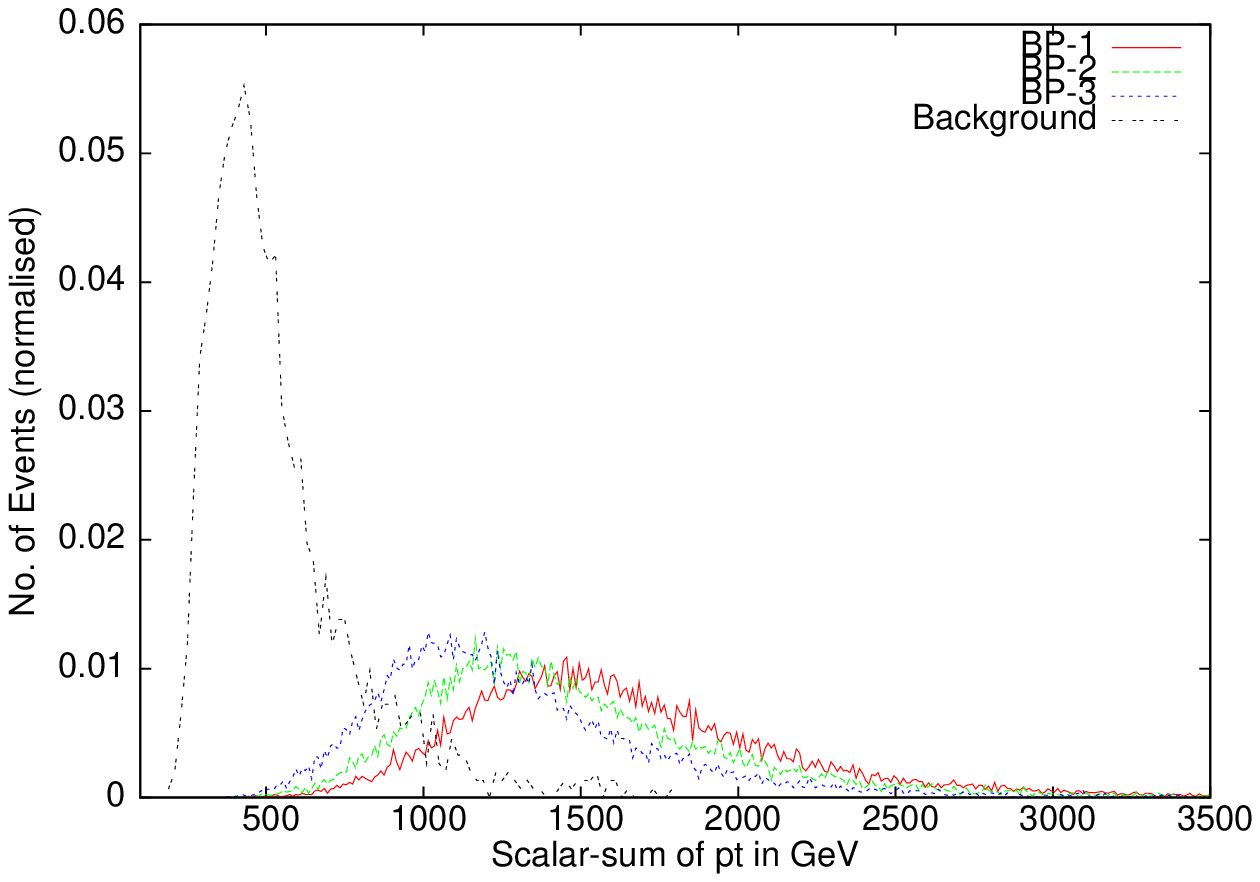,width=7.0cm,height=6.0cm,angle=-0}
\hskip 20pt \epsfig{file=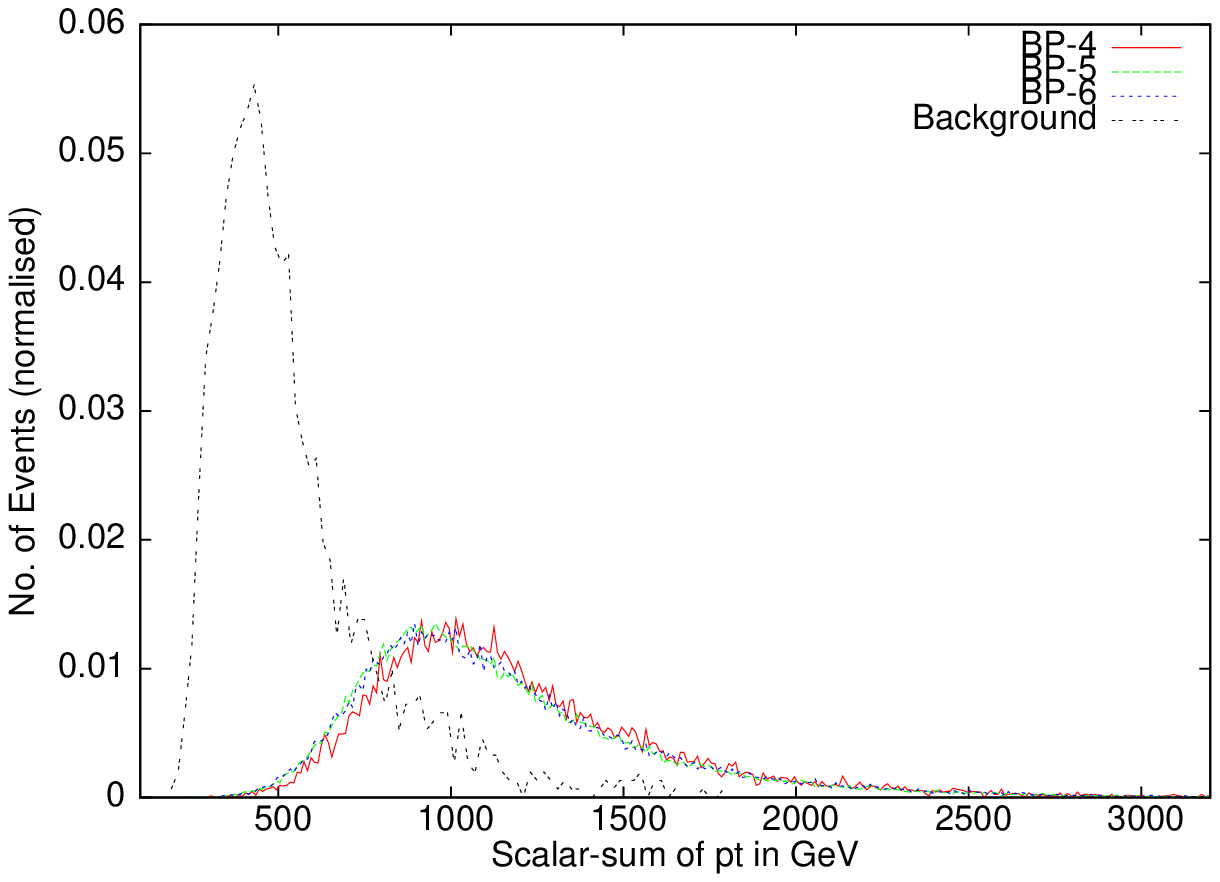,width=7.0cm,height=6.0cm,angle=-0}}
\caption{$\Sigma|\vec{p_T}|$ distribution (normalised to unity) for the signal and 
the background, for all benchmark points.} 
\end{center}
\end{figure}
\vspace{0.5cm}

In order to obtain the invariant mass of a tau-stau pair,
one needs to extract information on the mass of the stable
charged particle (the stau in our context). While standard 
techniques such as time delay measurement or the degree
of ionisation produced has been suggested in a number
of earlier works \cite{Hinchliffe:1998ys}, we extract mass 
information from an event-by-event
analysis which is reported in the next section. 
The efficiency for the reconstruction of staus has been 
taken to be the same as that of muons  
with $p_T>~10~GeV$ in the  pseudorapidity range $|\eta|<2.5$,
and is set at 90\% following \cite{TDR}.

{\noindent
{\bf $\tau$-jet tagging and $\tau$-reconstruction:} 

$\tau$-jet identification and $\tau$-reconstruction 
are necessary for both background reduction and mass 
reconstruction of the neutralinos. We have concentrated on 
hadronic decays of the $\tau$ in the one-prong channel. 
\footnote{We have not considered the leptonic decay of tau, as it is difficult to
identify lepton coming from tau decay to that coming from cascade decay of other
objects like heavy quarks or W's}. 
These are jet-like clusters in the 
calorimeter containing a relatively small number of charged 
and neutral hadrons. A $\tau$ decays  
hadronically about 65\% of the time, producing a $\tau$-jet.  
The momentum of such a jet in the plane transverse to the parent $\tau$
is small compared to the $\tau$-energy, so long as the $p_T$ 
of the $\tau$-jet is large compared to the $tau$ mass. In this
limit, hadronic $\tau$ decays produce narrow jets collinear with the
parent $\tau$. The neutrinos that carry missing $E_T$ also have
the same direction in this limit. 
This gives one a handle in reconstructing the $\tau$'s, if one
selects events where no other invisible particle is likely to 
be produced.

Given the masses of the SUSY particles in our benchmark scenarios, the
$\tau$'s produced out of neutralino decay are hard enough, so that one
can simulate $\tau$-decays in the collinear approximation described
above. A detailed discussion on the procedure followed for 
complete reconstruction of a pair of $\tau$'s 
is found in \cite{Rainwater:1998kj}. We have selected
hadronic jets with $E_T>50$ GeV as candidate products of $\tau$-decay.
A rather conservative non-tau jet rejection factor of 20
has been assumed,  while the identification efficiency of a
true tau-jet has been assumed to be 50\% following \cite{Coadou:tauidfic,CMS1}.\\

To describe the procedure in brief, one can fully reconstruct 
the $\tau$ by knowing $x_{\tau_{hi}}$ (i = 1,2), 
the fractions of the parent $\tau$-energy carried by each product jet.
The two unknowns can be solved from the two components of
the missing transverse momentum ($\vec{p_T}\miss$) of a
particular event.

If $p^{\mu}_{\tau_{i}}, p^{\mu}_{hi}$ are, respectively the
components of four-momentum of the parent $\tau$ and the collinear
jet produced from it (i = 1,2), then

\be
p^{\mu}_{hi}=x_{\tau_{hi}}~p^{\mu}_{\tau_{i}} ~~~~~~~~~~~~ (\mu=0,1,..,4)
\ee

(as $E_{\tau}\approx|\vec{p_{\tau}}|$, in the limit $m_{\tau}\r 0$)


and one can write

\be
\vec\sla{p_T} = ({1\over x_{\tau_{h1}}} - 1) \; \vec p_{h1} +
({1\over x_{\tau_{h2}}} - 1) \; \vec p_{h2} \; 
\ee

This yields two conditions for $x_{\tau_{hi}}$. Solving them, one
obtains the $\tau$ four-momenta as $p_{hi}/x_{\tau_{hi}}$.  In
practice, as will be discussed below, the recorded missing momentum,
$\vec\sla{p^{rec}_T}$, is different from the true one, namely,
$\vec\sla{p^{true}_T}$.  This error can lead to unphysical solutions
for the reconstructed $\tau$-momenta in some cases. Such a situation
often arises when the two taus are produced back-to-back.  This in
turn means that the $\tau$-neutrino's are also produced back-to-back in
the collinear approximation. This reduces the magnitude of
$\vec\sla{p_T}$, when errors due to mismeasurements can lead to unphysical
solutions. This is sometimes avoided by leaving out back-to-back
orientation of the two $\tau$-jet candidates, with some tolerance.  In our
analysis, a minimum value for $\sla{E_T}$ ($\approx40~GeV$) and
positivity of $x_{\tau_{hi}}$'s have been imposed as necessary
conditions, in order to minimise the number of unphysical solutions.
Besides, $p_T>50$ GeV for each $\tau$-jets ensures the validity of the
collinear approximation in $\tau$-decays. The $\tau$-identification
efficiency and the jet rejection factor are also better optimized
with this $p_T$-cut \cite{Coadou:tauidfic,CMS1}.

Of course, with a jet rejection factor of $1/20$, one cannot rule out
the possibility of QCD jets masquerading as $\tau$'s, in view of the
huge number of QCD events at the LHC. Such fakes constitute 
irreducible backgrounds to the $\tau$-reconstruction
procedure. However, as we shall see in the numerical results presented
in the next section, triggering on the rather strikingly spectacular 
properties of the quasi-stable stau-pair enables one to filter out the
genuine events in the majority of cases.

{\noindent
{\bf Reconstruction of $\vec\sla{p_T}$ :} 

It is evident from the above observations that the reconstruction of
$\vec\sla{p_T}$ is very crucial for our study. The reconstructed
$\vec\sla{p_T}$ differs considerably from true $\vec\sla{p_T}$, due to
several reasons. The true $\vec\sla{p^{true}_T}$ is related to the
experimentally reconstructed $\vec\sla{p^{rec}_T}$ by the following
relation

\be
\vec\sla{p^{rec}_T} =\vec\sla{p^{true}_T}+\vec\sla{p^{Forw}_T}+\vec\sla{p^{<0.5}_T}
\ee
\noindent
{where $\vec\sla{p^{Forw}_T}$ corresponds to the total transverse
momentum carried by the particles escaping detection in the range
$|\eta|>5$ and $\vec\sla{p^{<0.5}_T}$ corresponds to the total
transverse momentum carried by the particles in the range $|\eta|<5$
with $p_T<0.5$ GeV\footnote{The threshold is 0.5 GeV for CMS and 1 GeV
for ATLAS.}, which 
contributes to the true $\vec\sla{p_T}$. In addition to this,
mismeasurement of the transverse momenta for jets, leptons etc. alters 
the true $\vec\sla{p_T}$ by an sizable amount. This is due to the finite
resolution of detectors, and is handled in theoretical predictions
by smearing the energy/momentum of a particle through a Gaussian function.}

In our study, we have tried to reconstruct $\sla\vec{p_T}$, taking into
account all the above issues. The missing transverse momentum
in any event is defined as

\hspace{3.0cm} $\vec{\sla{p_T}}=-\Sigma\vec{p}^{visible}_T$\\
\noindent
{where the $\Sigma\vec{p}^{visible}_T$ consists of isolated
leptons/photons/jets and also those objects which do not belong to any
of these components but are detected at the detector, constituting the
so called `soft part' or the `unclustered component' of the visible
momentum. We describe below the various components of the visible
$\vec{p_T}$, and their respective resolution functions.}

{\noindent
{\bf Resolution effects:} 

Among the finite resolution effects of the detector, taken into account
in our analysis, most important are the finite resolutions of
the`electromagnetic and hadron calorimeters, and the muon track
resolution. Since the kind of final state we are dealing with does not
require any isolated electrons/photons, we have not considered
electron or photon resolution. The electrons/photons which are not
isolated but have $E_T\geq 10$ GeV and $|\eta|< 5$ have
been considered as jets and their resolution has been parametrised
according to that of jets. Jets have been defined within a cone of 
$\Delta R=0.4$ and $E_T\geq 20 GeV$ using the PYCELL fixed cone jet 
formation algorithm in PYTHIA. Since the staus are long lived and live a
charged track in the muon chamber, their smearing criteria have been
taken to be the same as those of isolated muons. Though one can
describe the resolution of the track of staus and muons by different
resolution functions (as $m_{\stau}>> m_{\mu}$), one does not envision
any significant deviation in the prediction of events via such
difference. Therefore, in the absence of any clear guidelines, we have
treated them on equal footing, as far as the Gaussian smearing
function is concerned. The tracks which shows up in the muon chamber,
but are not isolated, having $E_T>10$GeV and $|\eta|< 2.5$, have been
considered as jets and smeared accordingly. 
All the particles (electron, photon, muon, and stau) with $0.5<
E_T< 10$GeV and $|\eta|< 5$ (for muon or muon-like tracks,
$|\eta|< 2.5$), or hadrons with $0.5< E_T< 20$GeV and $|\eta|<
5$, which constitute `hits' in the detector, are considered as {\it soft
or unclustered components}. Their resolution function have been considered
separately. We present below the different parametrisation of the
different component of the final state, assuming the smearing to be 
Gaussian in nature. \footnote {Although a "double Gaussian" smearing
is followed in more realistic detector simulations, our illustrative
study is now substantially affected by such considerations.} 
\\
\begin{itemize}

\item {\bf Jet energy resolution :}

\be
\sigma(E)/E=a/\sqrt{E}\oplus b\oplus c/E
\ee

where\\
$~~~~~~~~$ a= 0.7 [GeV$^{1/2}$], $~~$    b= 0.08 \& $~$    c= 0.009 [GeV] $~~~$  for $|\eta|< 1.5$ \\     
$~~~~~~~~~~$= 1 $~~~~~~~~~~~~~~~~~~~~$    = 0.1  $~~~~~~~$ = 0.009 $~~~~~~~~~$ $1.5< |\eta|< 5$
        
\item {\bf Muon/Stau $p_T$ resolution :}

\begin{eqnarray}
\sigma(p_T)/p_T&=&a  ~~~~~~~~~~~~~~~~~~~~~~~{\rm if} ~ p_T<\xi\\
&=&a+b\log(p_T/\xi) ~~~~~ {\rm if}~ p_T>\xi
\end{eqnarray}

where\\
$~~~~~~~~$ a= 0.008, $~~$    b= 0.037 \& $\xi=100$ [GeV]$~~~~$ for $|\eta|< 1.5$ \\     
$~~~~~~~~~~$= 0.02 $~~~~~~$    = 0.05  $~~~~$ $~\xi=100~~~~~~~~$ $1.5< |\eta|<2.5$

\item {\bf Soft component resolution :}

\be
\sigma(E_T)=\a\sqrt{\Sigma_{i}E^{(soft)i}_T}
\ee

with $\a\approx0.55$. One should keep in mind that the x and y
component 
of $E^{soft}_T$ need to be 
smeared independently and by the same quantity.

\end{itemize}}


It is of great importance to ensure that the stable $\tilde{\tau}$
leaving a track in the muon chamber is not faked by an actual muon
arising from a standard model process. As has been mentioned in
section 1, we have found certain kinematic prescriptions to be 
effective as well as reliable in this respect. In order to see this
clearly, we present the $p_T$-distributions of the harder muon and
the $\tilde{\tau}$-track in Figure 1. The $\tilde{\tau}$-$p_T$ clearly
shows a harder distribution, owing to the fact that the stau takes
away the lion's share of the  $p_T$ possessed by the parent
neutralino. Another useful discriminator is the scalar sum of
transverse momenta of all detected particles (jets, leptons and unclustered
components). The distribution in $\Sigma|\vec{p_T}|$, defined in the above
manner, displays a marked distinction for the signal events, as shown
in Figure 2. The cuts chosen in Table 3 have been guided by both of
the above considerations.  They have been applied for all 
the benchmark points, as also for the background calculation. 

\renewcommand{\baselinestretch}{1.1}\selectfont
\vspace{-0.5cm} 
\begin{table}[htb]
\begin{center}
\footnotesize
\begin{center}
\begin{tabular}{ |l||c| }
\hline
\hline
\multicolumn{2}{|c|} { Cuts } \\
\hline
\hline
               & $ p_{T}^{lep, stau} > 10$~GeV\\ 
               & $ p_{T}^{hardest-jet} > 75$~GeV  \\
               & $ p_{T}^{other-jets} > 50$~GeV \\
 ~~Basic Cuts  & $ 40~GeV<\sla{E_T} < 150~ GeV$ \\
               & $|\eta| < 2.5$  for Leptons, Jets \& Stau \\
               & $\Delta R_{ll}>0.2,~ \Delta R_{lj}>0.4$\\
               & $\Delta R_{\stau l}>0.2, ~\Delta R_{\stau j}>0.4$\\
               & $ \Delta R_{jj} >0.7$ \\
\hline

  Cuts for     & $p_T^{iso~charg~track}> 100~GeV$ \\
 Background    & $\Sigma |\vec{p_T}| > 1 ~TeV$ \\
 Elimination   &  \\

\hline

Invariant mass difference  & $|M^{pair1}_{\stau\tau}-M^{pair2}_{\stau\tau}|< 50~GeV$ \\
 of two nearby pairs & \\
\hline
\hline
\end{tabular}
\caption{\small  \it Cuts applied for event selection, background elimination
and neutralino reconstruction.}

\end{center}
\end{center}
\vspace{-0.5cm}
\label{tab:Cuts}
\end{table}

\vspace{1.0cm}

\section {Numerical results and neutralino reconstruction:}

\subsection {The reconstruction strategy}

Having obtained the $\tau$ four-momenta, the neutralinos can be
reconstructed, once we obtain the 
energy of the $\tilde{\tau}$'s whose
three-momenta are already known from the curvature of the tracks
in the muon chamber. 
For this, one needs to know the  $\tilde{\tau}$-mass. In addition,
the requirements are, of course, sufficient statistics, minimisation
of errors due to QCD jets faking  the $\tau$'s, and the
suppression of combinatorial backgrounds. For the first of these,
we have presented our numerical results uniformly for an integrated 
luminosity of 300 $fb^{-1}$, although some of our benchmark points
requires much less luminosity for effective reconstruction. We have
already remarked on the possibility of reducing the faking of $\tau$'s. 
As we shall show below, a systematic  procedure can also be adopted
for minimising combinatorial backgrounds to the reconstruction
of neutralinos. The primary step in this is to combine each such
$\tau$ with one of the two hardest tracks in the muon chamber, which
satisfy the cuts listed in Table 3. A particular $\tau$ is combined with
a heavy track of opposite charge. However, since neutralinos 
are Majorana fermions, producing pairs of $\tau^+ \tilde{\tau}^-$
and $\tau^- \tilde{\tau}^+$ with equal probability, this is not enough
to avoid the combinatorial backgrounds. Therefore, out of the
two $\tau$'s and two heavy tracks, we select those pairs which give 
the closer spaced invariant masses, with
$|M^{pair1}_{\stau\tau}-M^{pair2}_{\stau\tau}|<50 ~GeV$.  The number of signal and 
background events, after the successive application of cuts are listed in Table 4.\\

\renewcommand{\baselinestretch}{1.1}\selectfont
\begin{table}[htb]
\begin{center}
\footnotesize 
\begin{center}
\begin{tabular}{||l||r r r r r r||r|r|r||r||}
\hline
\hline
 {~~~CUTS} & \multicolumn{6}{c||}{SIGNAL } & \multicolumn{4}{c||}{BACKGROUND }\\ 
 &BP1 & BP2 & BP3 &BP4 &BP5 &BP6 &$t\bar{t}$ & {ZZ} & {ZH} & total \\
\hline
  Basic Cuts   &1765 &4143 &11726 &20889 &28864 &15439 & 4129 & 45 & 6 & 4180 \\
$+P_T$ Cut       &1588 &3631 &9471 &15526 &20282 &9920 & 210 & 3 & 1 & 214 \\
$+\Sigma{|P_T|}$ Cut &1442 &3076 &6777 &9538 &11266 &5724 & 63 & 0 & 0 &  63  \\
$+|M^{pair1}_{\stau\tau}-M^{pair2}_{\stau\tau}|$ Cut  
                         &408 &887 &1622 &2004 &2244 &858 & 6 & 0 & 0 &  6 \\
\hline 
\hline
\end{tabular}
\caption{\small \it {Number of signal and background events for
    $2\tau_j+2\stau$ (charged-track)+$E_{T}\miss+X$ final state with
    an integrated luminosity of $300~fb^{-1}$, considering all SUSY
    processes. The standard model Higgs mass is taken to be 120 GeV in
    background calculation.}}
\label{tab. Events}
\end{center}
\end{center}
\end{table}

This finally brings us to the all-important issue of knowing the
stau-mass. The $\stau$-mass can be reconstructed from the information 
on time delay ($\Delta t$) between the inner tracker and the outer muon
system and the measured three-momentum of the charged track
\cite{Hinchliffe:1998ys}. The accuracy of this method
depends on the accurate determination of $\Delta t$, which can be limited
when the particles are highly boosted. We have followed a somewhat
different approach to find the actual mass of the particle associated
with the charged track. We have found this method effective when
both pairs of  $\tau \tilde{\tau}$  come from $\chi^0_1 \chi^0_1$
or  $\chi^0_2 \chi^0_2$. \\

{\bf Solving  for the $\stau$-mass:} The actual $\stau$-mass can be
extracted by demanding that the invariant mass of the two correct
$\stau\tau$-pairs are equal, which yields an equation involving one
unknown, namely, $m_{\stau}$:

\be
\sqrt{m^2_{\stau}+|\vec{p_{\stau_1}}|^2}.E_{\tau_1}-\sqrt{m^2_{\stau}+|\vec{p_{\stau_2}}|^2}.E_{\tau_2}
=\vec{p_{\stau_1}}.\vec{p_{\tau_1}}-\vec{p_{\stau_2}}.\vec{p_{\tau_2}}
\ee
\noindent{
where the variables have their usual meanings and are
experimentally measurable event-by-event. ($\tilde{\tau}_{1,2}$ here
denote the lighter $\stau$s on two sides of the cascade, and not
the two $\stau$ mass eigenstates.)}\\

The right combination is assumed to be selected whenever the difference
between $|M^{pair1}_{\stau\tau}- M^{pair2}_{\stau\tau}|$ is minimum
and differ by not more than 50 GeV, as mentioned earlier.
It should be noted that the unambiguous identification of the right
$\tau\stau$-pairs which come from decays of two neutralinos
($\chi^0_1\chi^0_1$ or $\chi^0_2\chi^0_2$) does not depend
on the actual stau mass. Thus we can use a `seed value' of the stau
mass as input to the above equation, in identifying the right $\tau\stau$
combinations. We have used a seed value of 
$m_{\stau}\approx 100~GeV$ (motivated
by the LEP limit on $m_{\stau}$). The SM background has already
been suppressed by demanding the $p_T$ of each charged track to be
greater than 100 GeV, together with $\Sigma|\vec{p_T}| > 1~TeV$.\\

Once the right pairs are chosen using the seed value of the
$\stau$-mass, we need not use that value any more, and instead solve
equation 16 which is quadratic in $m^2_{\stau}$. We have kept only
those events in which at least one positive solution for $m^2_{\stau}$
exists. When both roots of the equation are positive, the higher value
is always found to be beyond the reach of the maximum centre-of-mass
energy available for the process. Hence we have considered the
solution corresponding to the lower value of the root. The
distribution of the solutions thus obtained has a peak around the
actual $\stau$-mass. The $\stau$-track four-momenta are
completely constructed, using this peak value as the actual mass
of the $\stau$-NLSP (see Figure 3). The fact that these peaks faithfully
yield the $\stau$-mass (see Table-1), makes it unnecessary to extract this 
mass from 'fits'.\\

\begin{figure}[tbhp]
\begin{center}
\centerline{\epsfig{file=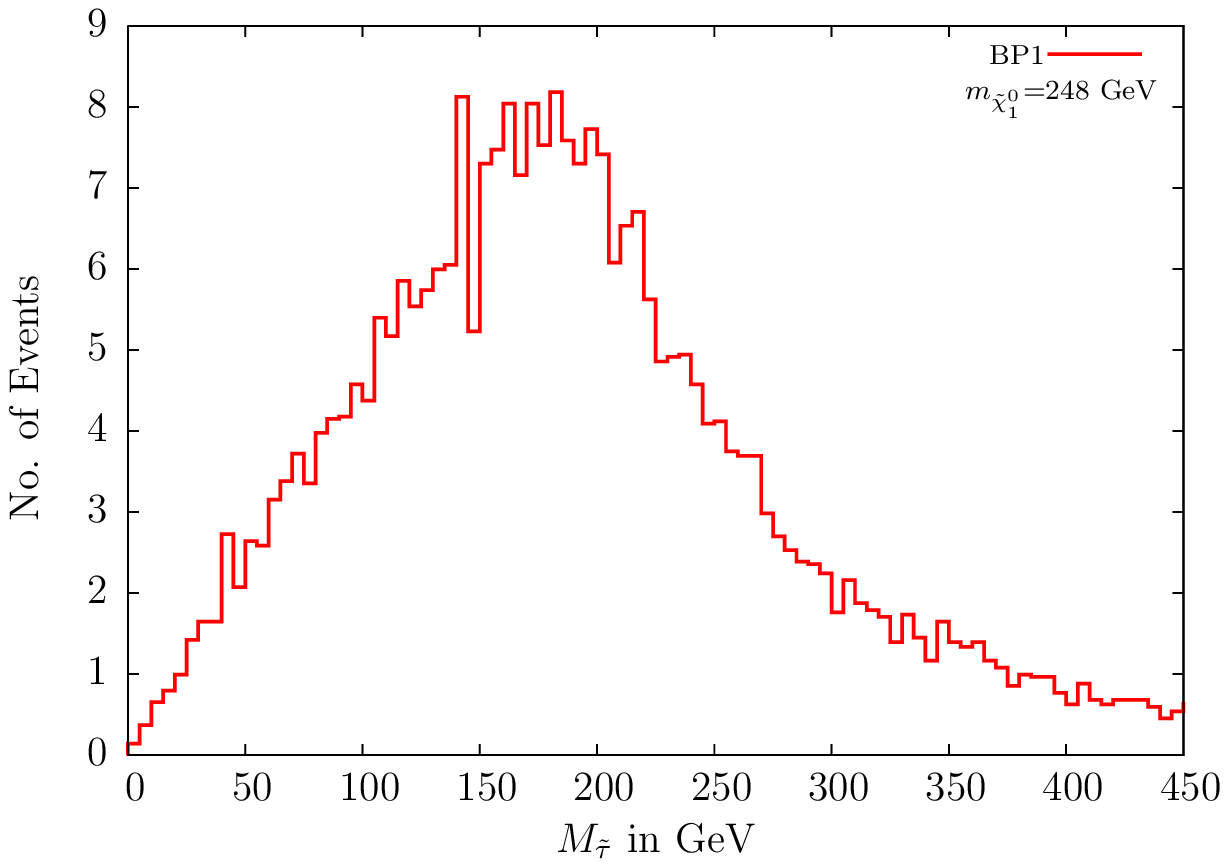,width=7.0cm,height=6.0cm,angle=-0}
\hskip 20pt \epsfig{file=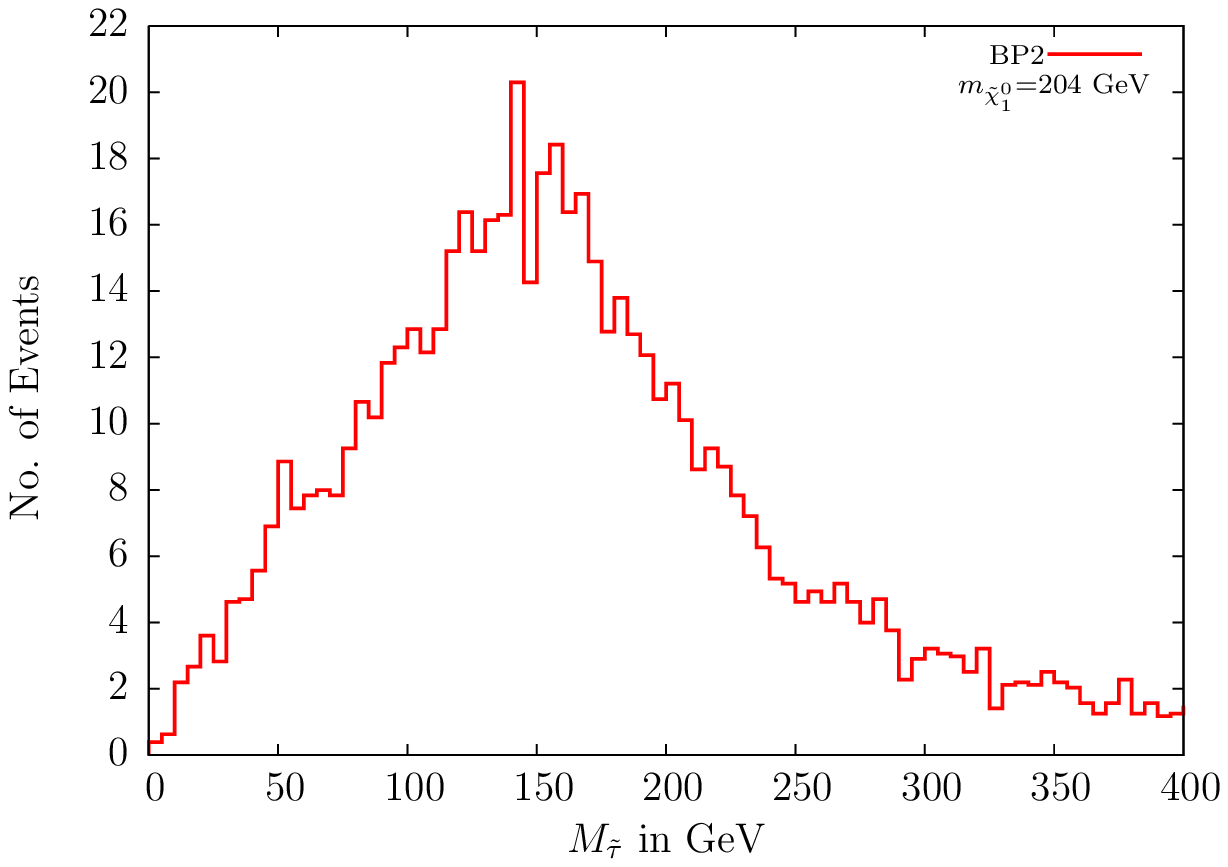,width=7.0cm,height=6.0cm,angle=-0}}
\vskip 10pt
\centerline{\epsfig{file=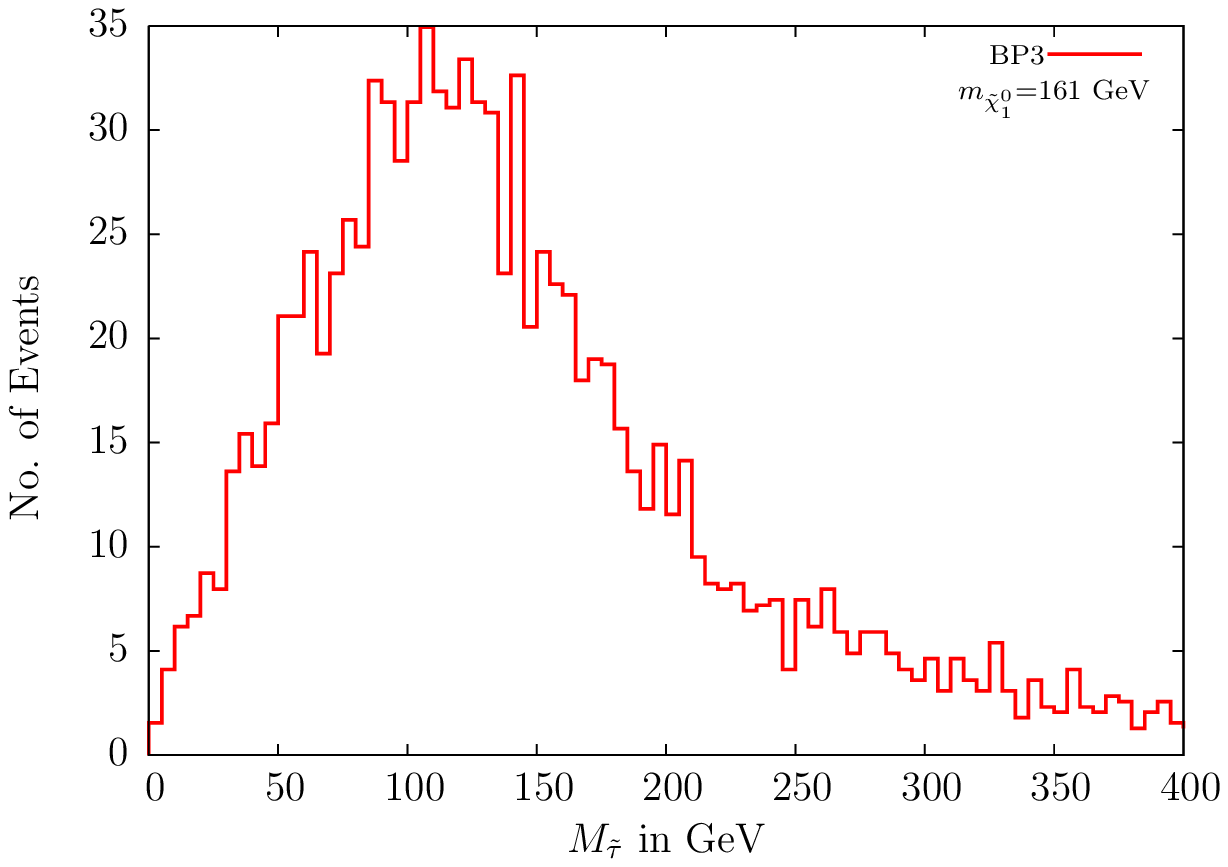,width=7.0cm,height=6.0cm,angle=-0}
\hskip 20pt \epsfig{file=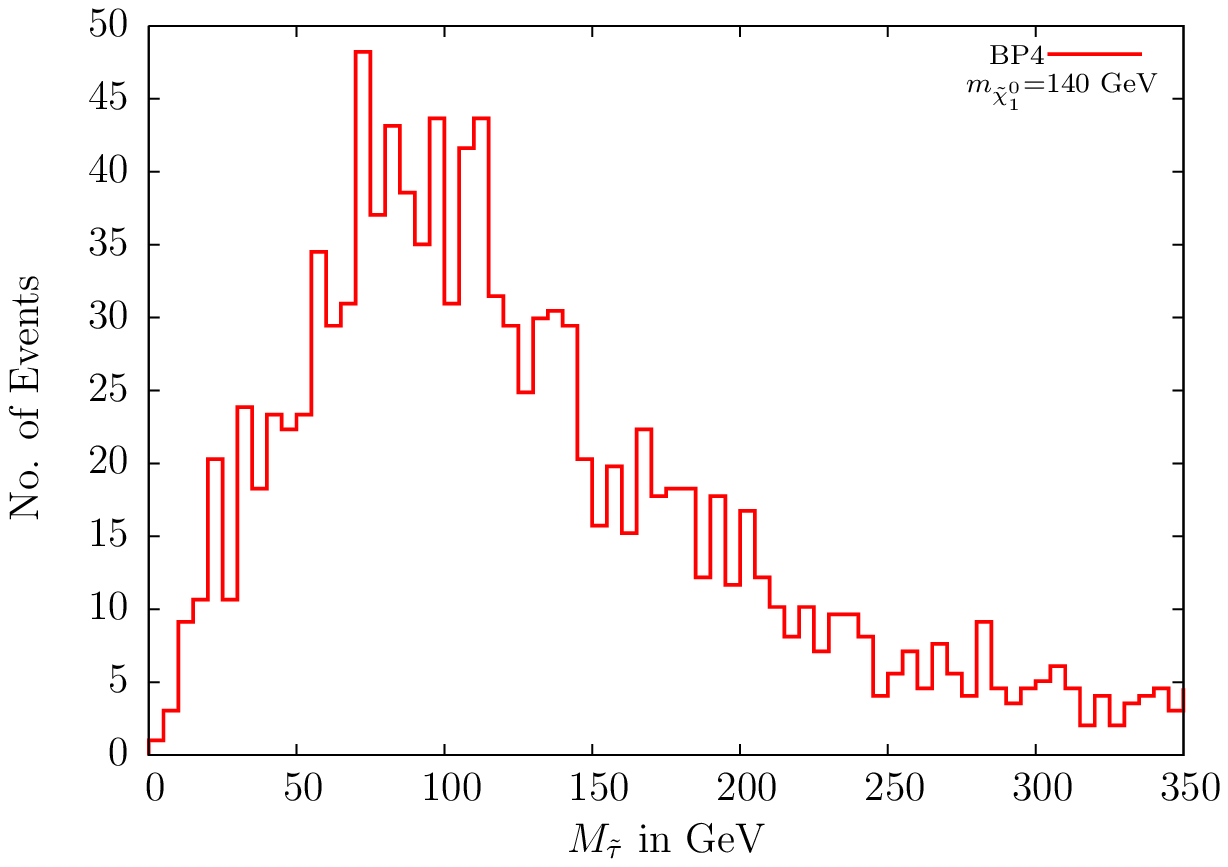,width=7.0cm,height=6.0cm,angle=-0}}
\vskip 10pt
\centerline{\epsfig{file=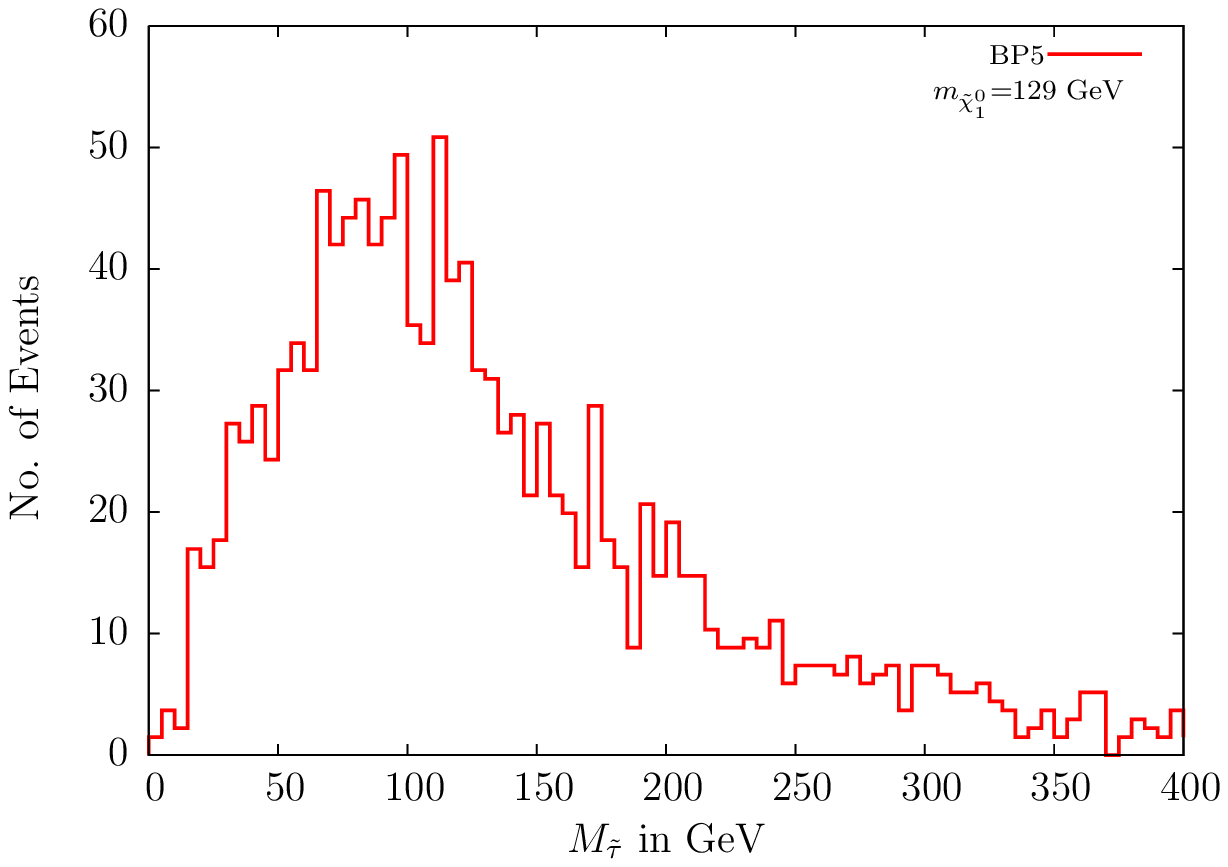,width=7.0cm,height=6.0cm,angle=-0}
\hskip 20pt \epsfig{file=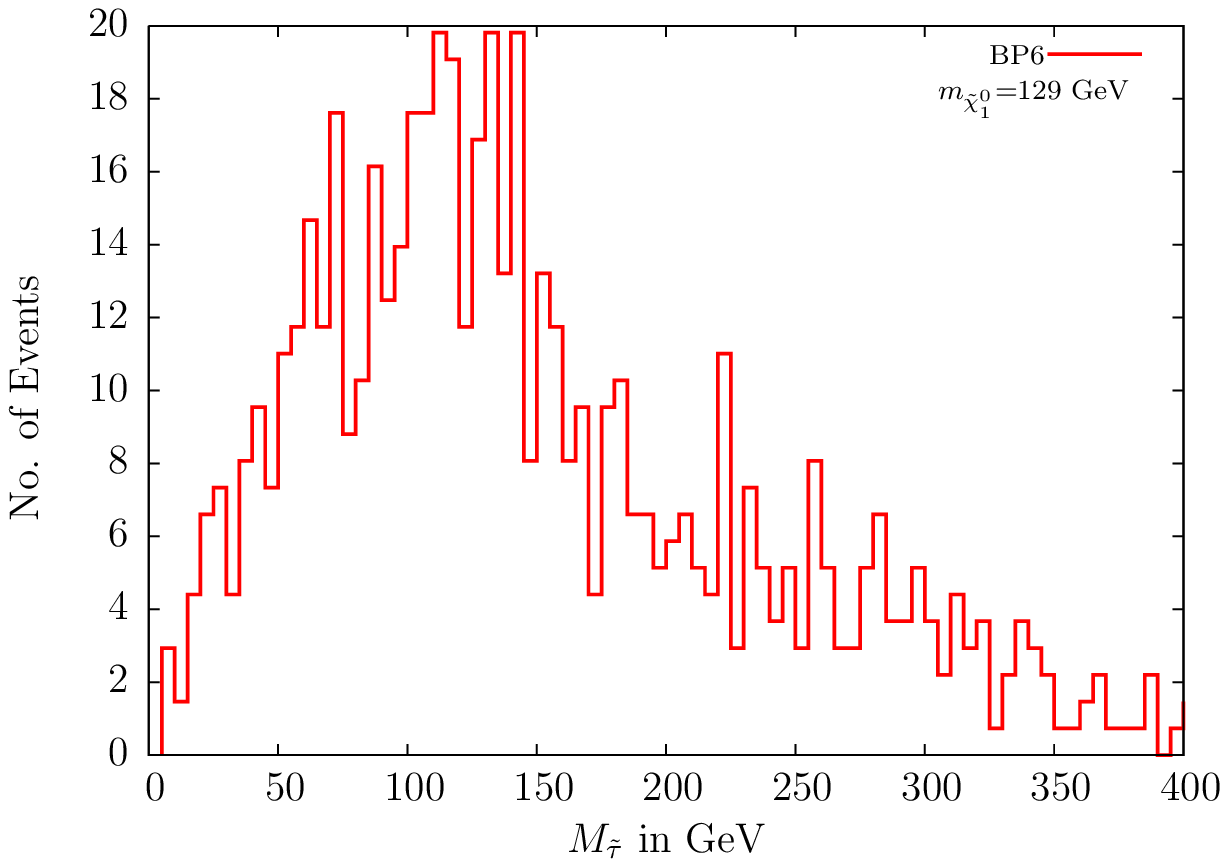,width=7.0cm,height=6.0cm,angle=-0}}

\caption{The $\stau$ mass peak as obtained from eventwise reconstruction as
described in the text, for all the benchmark points at luminosity 300 $fb^{-1}$.} 
\end{center}
\end{figure}

\begin{figure}[tbhp]
\begin{center}
\centerline{\epsfig{file=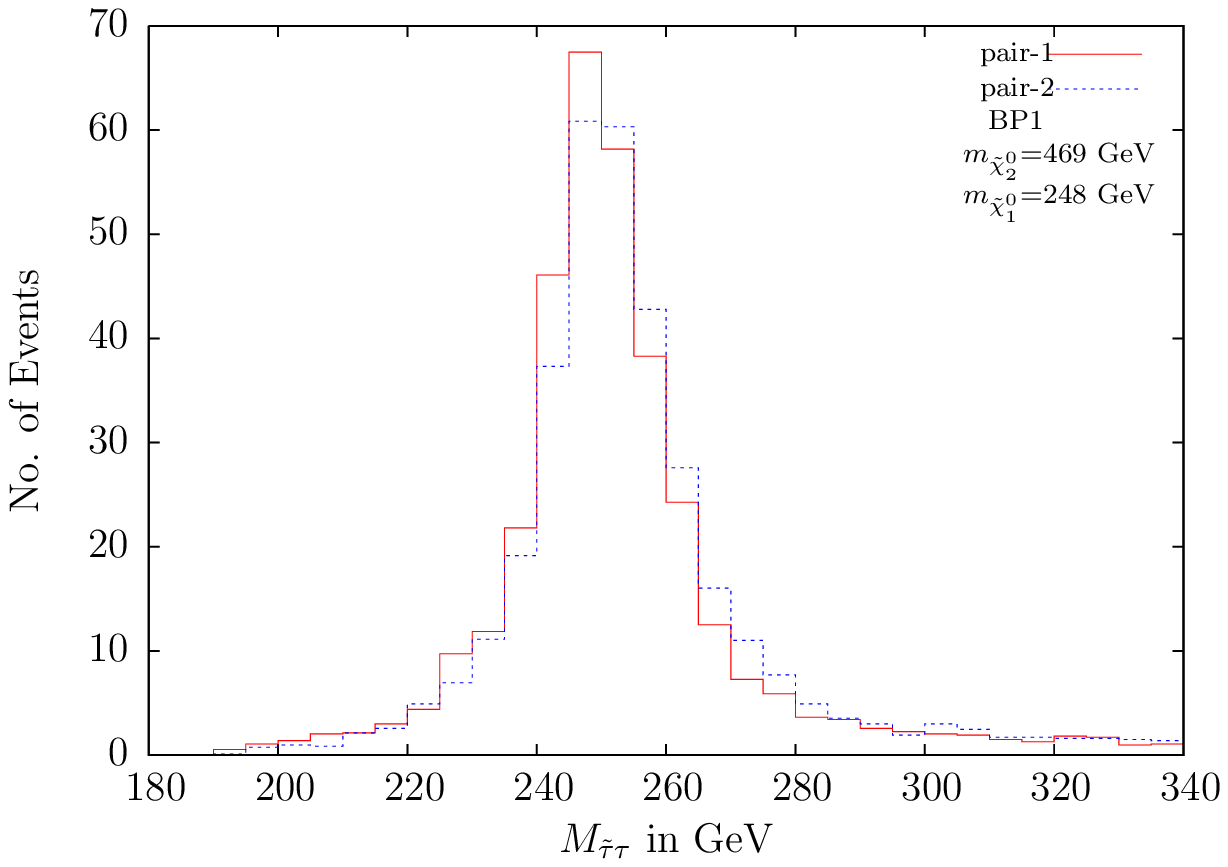,width=7.0cm,height=6.0cm,angle=-0}
\hskip 20pt \epsfig{file=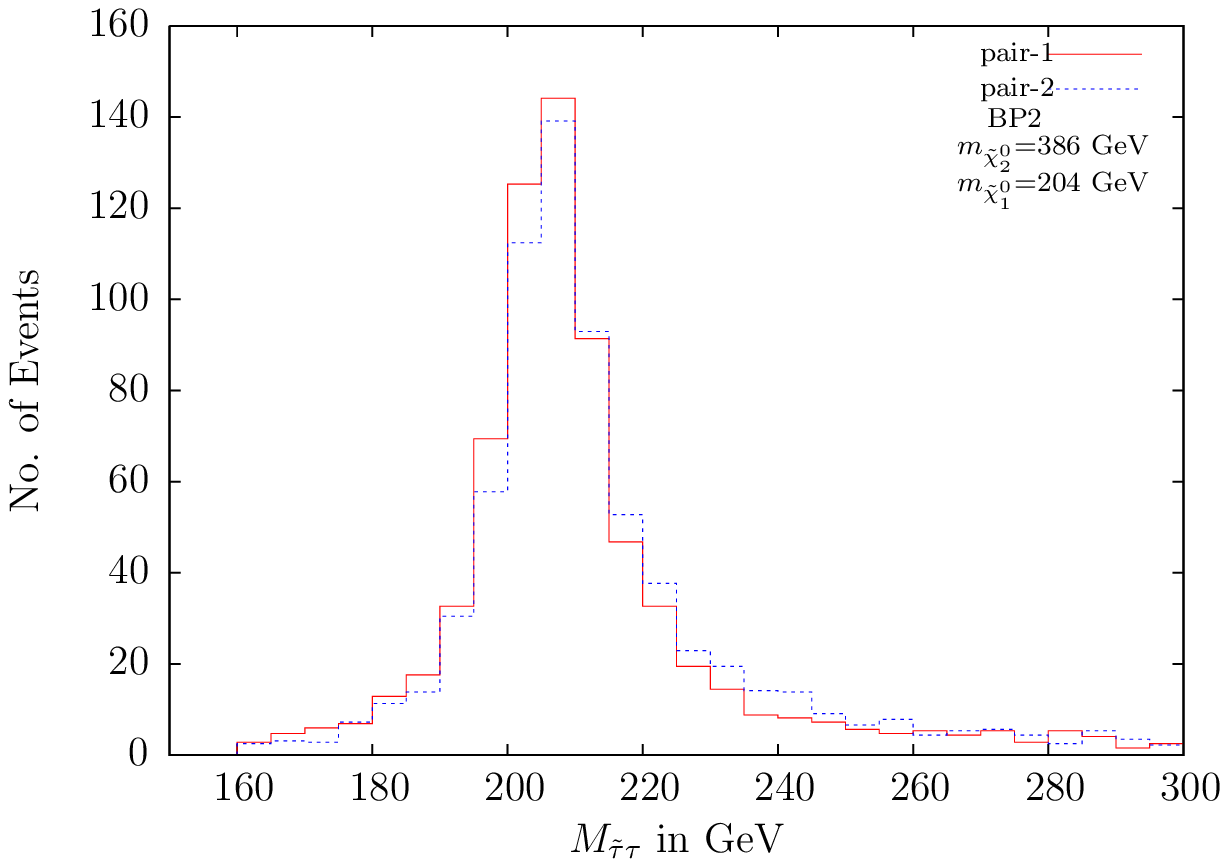,width=7.0cm,height=6.0cm,angle=-0}}
\vskip 10pt
\centerline{\epsfig{file=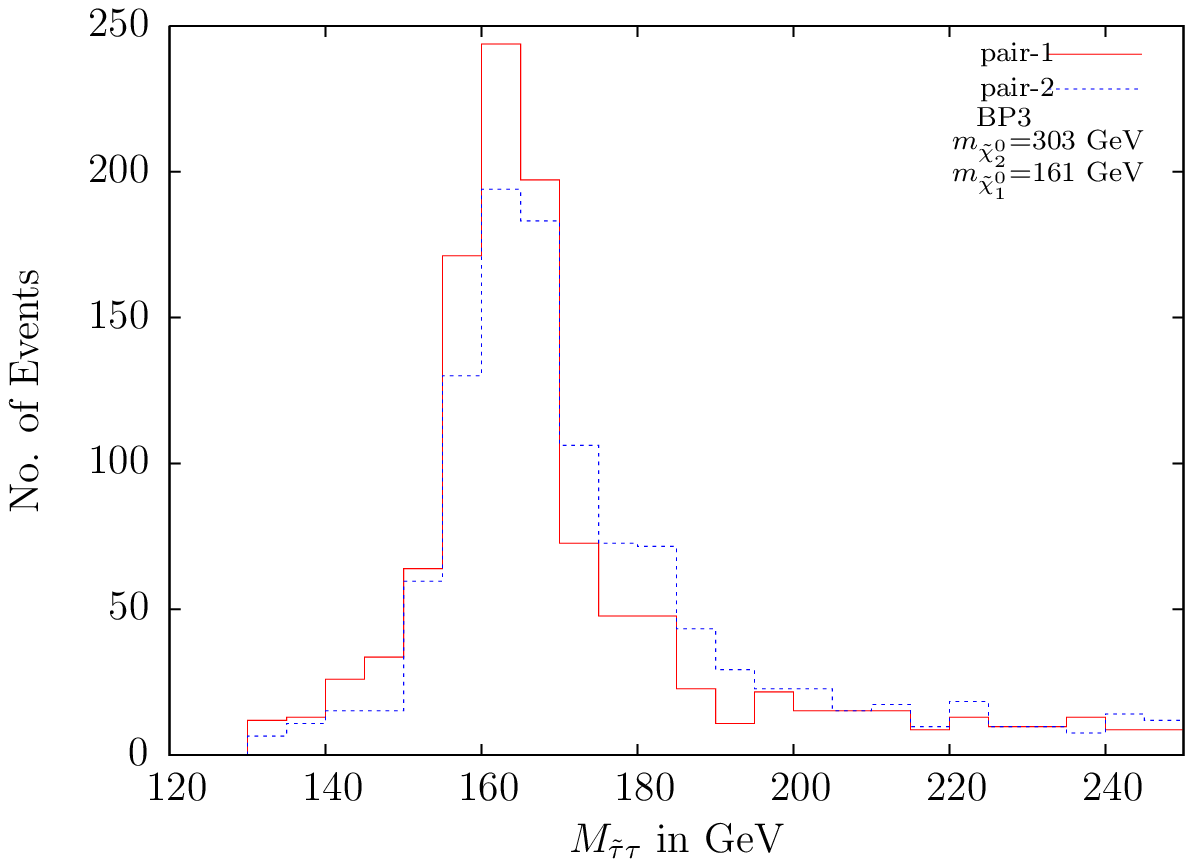,width=7.0cm,height=6.0cm,angle=-0}
\hskip 20pt \epsfig{file=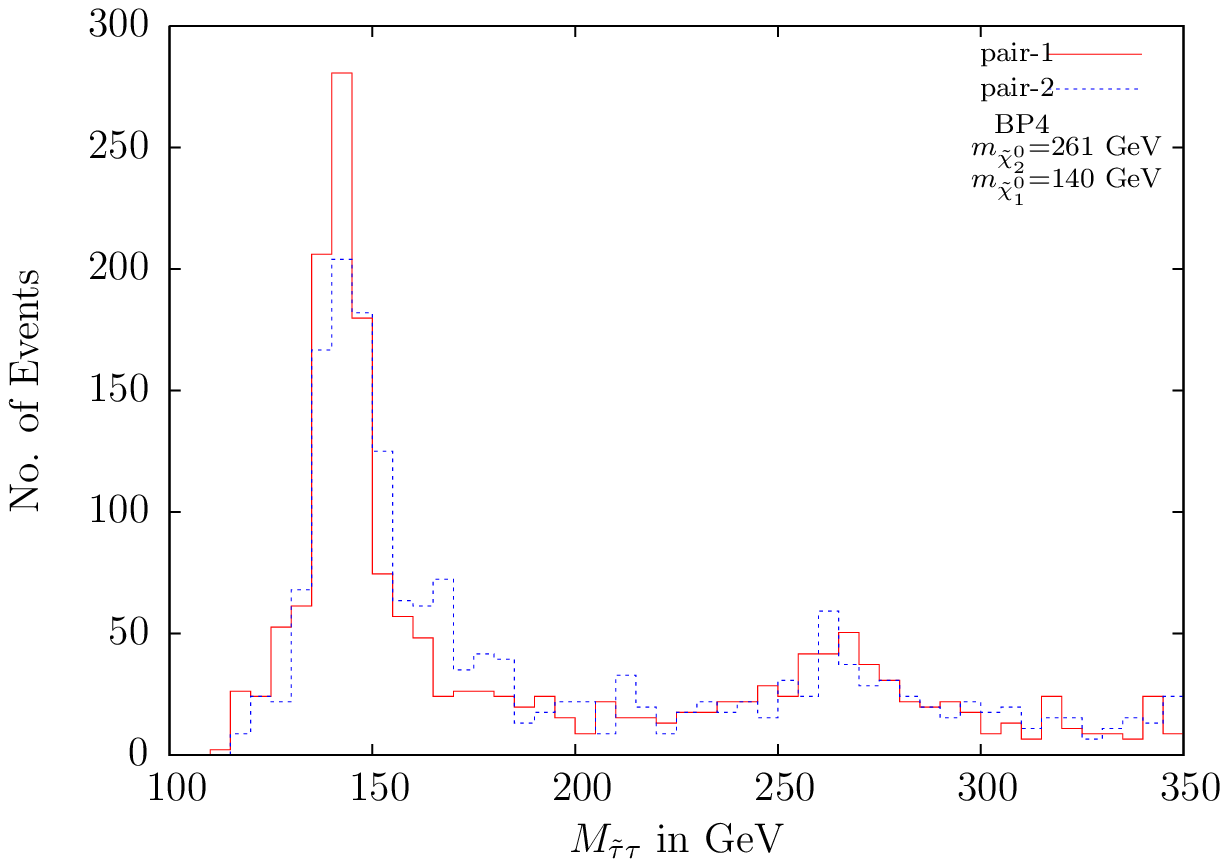,width=7.0cm,height=6.0cm,angle=-0}}
\vskip 10pt
\centerline{\epsfig{file=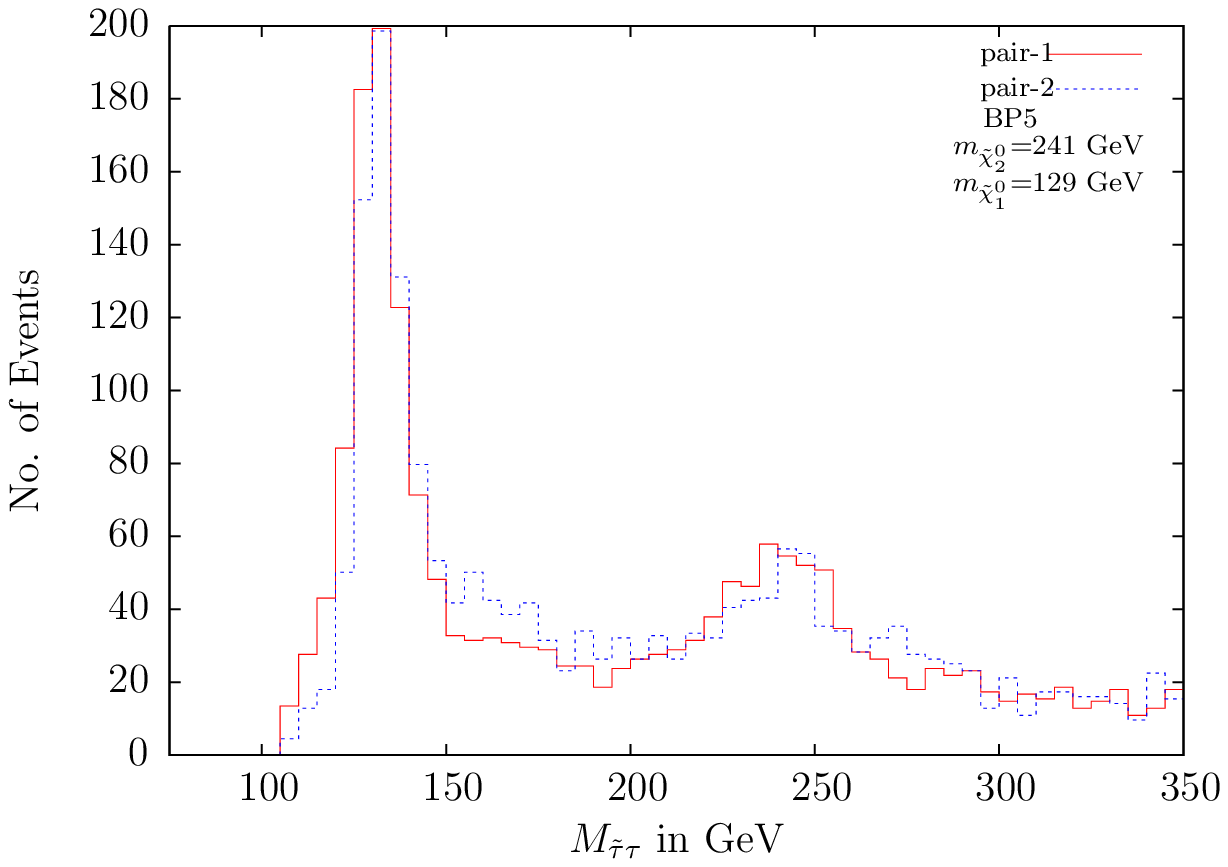,width=7.0cm,height=6.0cm,angle=-0}
\hskip 20pt \epsfig{file=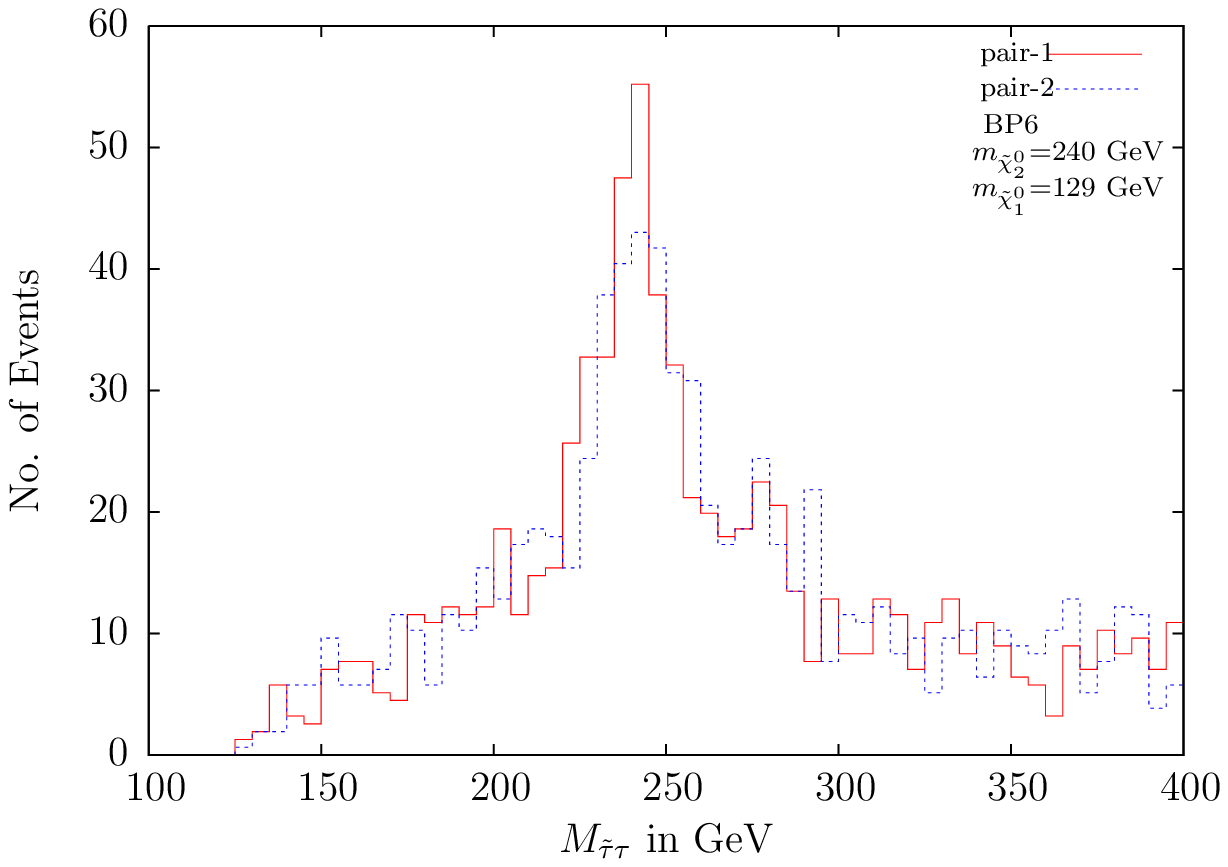,width=7.0cm,height=6.0cm,angle=-0}}
\caption{$M_{\stau\tau}$ distribution for all the benchmark points at luminosity 300 $fb^{-1}$. BP1, BP2 and
BP3 show only the $\chi^0_1$ peak. Both the $\chi^0_1$ and $\chi^0_2$ peaks are visible
for BP4 and BP5, while BP6 displays only the $\chi^0_2$ peak.} 
\end{center}
\end{figure}

This sets the stage fully for the reconstruction of neutralinos,
the results of which are shown in Figure 4.
For BP1, BP2 and BP3 one can see that there is only one peak which
corresponds to the $\chi^0_1$.
This is because the $\chi^0_2$ production rate in cascade
is relatively small for these points. For 
BP4 and BP5, on the other hand,  we have distinct peaks for both $\chi^0_1$ and
$\chi^0_2$. At BP6, however,  we only have the  $\chi^0_2$ peak. 
This is due to the small mass
splitting between $\chi^0_1$ ($M_{\chi^0_1}=129~GeV$) and
$\stau$ ($M_{\stau}=124~GeV$), which softens the tau (jet) arising
from its decay, preventing it from passing the requisite hardness cuts. 
On the whole, it is clear from Figure 4 (comparing the peaks
with the input values of the neutralino masses) that, in spite of
adulteration by QCD jets that fake the $\tau$'s, our event selection criteria 
can lead to faithful reconstruction of neutralino masses.

One may still like to know whether a neutralino reconstructed in this manner
is the  $\chi^0_1$ or the $\chi^0_2$, when only one peak is
visible. This requires further information on the SUSY spectrum.
For example, the information on the gluino mass, extracted from
the effective mass (defined as $(\Sigma|\vec{p}_T|~+~\sla{E}_T)$)
distribution, may enable one to distinguish between the 
$\chi^0_1$ and the $\chi^0_2$, once gaugino mass unification at 
high scale is assumed.

\subsection {LHC reach in the $m_0-M_{1/2}$ -plane:}

We have also identified the region in the $m_0-M_{1/2}$ plane,
where at least one of the two lightest neutralinos can be reconstructed.
For this, we have scanned over the region of the $m_0-M_{1/2}$ plane 
using the  spectrum generator SuSpect (v 2.34) \cite{SUSPECT} which leads 
to a $\stau$ LSP in a usual mSUGRA scenario without the 
right handed sneutrino \cite{Gladyshev:2006rx}. Results of this scan are shown in 
Figure 5. The coloured (shaded) areas are
consistent with all the low energy constraints like $b\r s\gamma$,
$B_s\r \mu^+\mu^-$, $\Delta(g_{\mu}-2)$ and the LEP limits on the low
energy spectrum. The value of $tan\beta$ has been fixed at 30,
 and $A_0$ = 100 has been chosen.

\begin{figure}[h]
\begin{center}
\epsfig{file=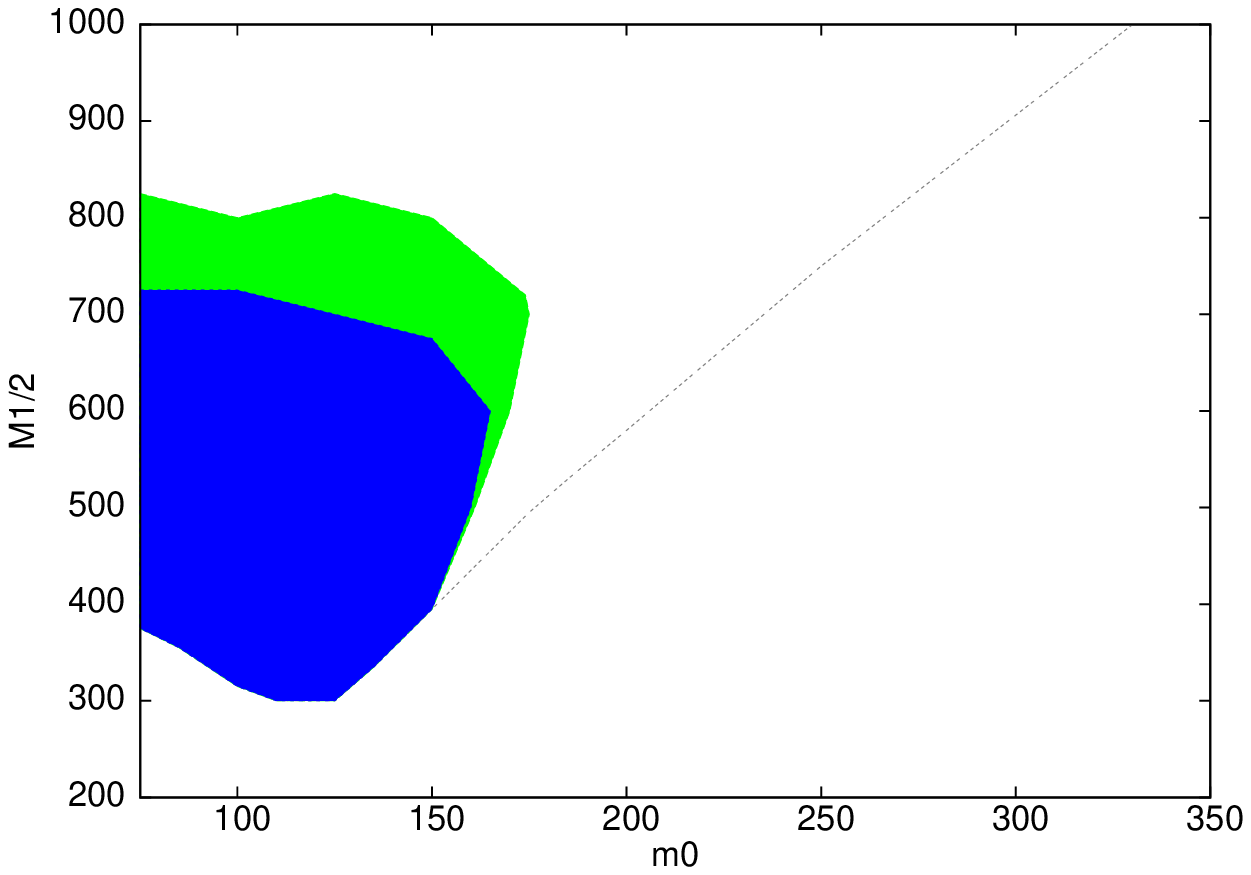,width=10.0cm,height=7.0cm,angle=-0}
\caption{The region  in the $m_0-M_{1/2}$ plane, where it is possible to reconstruct 
at least one of the neutralinos at the LHC, with $tan\beta=30$ and $A_0=100$. In the blue (dark shade) 
region, at least 100 events are predicted in the vicinity of the peak. The additional available region
where 50 events in the vicinity of the peak are assumed to suffice for reconstruction, is marked 
in green (light shade). The entire region above the dashed line indicates the scenario where 
one has a $\td{\nu_R}$-LSP and a $\stau$-NLSP.}
\end{center}
\end{figure}

The regions where reconstruction is possible have been determined using
the following criteria:

\begin{itemize}

\item In the parameter space, we have not gone into regions where the gluino 
mass exceeds $\approx 2 ~TeV$.

\item The number of events satisfying 
$|M_{\stau\tau}-M_{peak}|<0.1.M_{peak}$ at an 
integrated luminosity of 300 $fb^{-1}$ must be greater
than a specific number in order that the peak is said
to be reconstructed. One obtains the blue (dark shade) region if this number 
is set at 100. If the peak can be constructed from more than 50 events,
the additional region, marked in green (light shade), becomes
allowed. 
\end{itemize}

\section {Summary and conclusions}

We have considered a SUGRA scenario, with universal scalar and 
gaugino masses, where a right-chiral neutrino superfield exists for
each family. We have identified several benchmark points in the
parameter space of such a theory, where a right-sneutrino is the LSP,
and a $\stau$ mass eigenstate is the NLSP. The $\stau$, stable on the 
distance scale of the detectors, leaves a charged track in the muon chamber, 
which is the characteristic feature of SUSY signals in this scenario.
We use this feature to reconstruct neutralinos in the $\tau\stau$ channel.
For this, we use the collinear approximation to obtain the four-momentum
of the $\tau$, and suggest a number of event selection criteria to
reduce backgrounds, including combinatorial ones. We also suggest
that the $\stau$ mass may be extracted by solving the equation 
encapsulating the equality of invariant masses of two  $\tau\stau$ pairs
in each event. We find that at least one of the two lightest neutralinos 
can be thus reconstructed clearly  over 
a rather large region in the $m_0-M_{1/2}$ plane, following our
specified criteria.\\

{\bf Acknowledgment:} We thank Bruce Mellado for
his valuable suggestions on several technical points, and Subhaditya
Bhattacharya for giving important comments on the manuscript. We also
thank Priyotosh Bandyopadhyay, Asesh K. Datta, 
Nishita Desai and Sudhir K. Gupta  for helpful discussions. 
This work was partially supported by funding available 
from the Department of Atomic Energy, Government of India for the 
Regional Centre for Accelerator-based Particle Physics, 
Harish-Chandra Research Institute. Computational work for this study was
partially carried out at the cluster computing facility of
Harish-Chandra Research Institute ({\tt http:/$\!$/cluster.mri.ernet.in}).


\end{document}